\documentclass[10pt, conference, compsocconf, a4paper]{IEEEtran}
\DeclareFixedFont{\auacc}{OT1}{phv}{b}{it}{18}   % Patch by Gerry 3/21/07
\DeclareFixedFont{\newauacc}{OT1}{ptm}{b}{rm}{12}   % Patch by Gerry 3/21/07
\usepackage{hyperref}
\usepackage{cite}
\usepackage{graphicx}
\usepackage[tight]{subfigure}
\usepackage{breakurl}
\usepackage[amsthm]{ntheorem}
\usepackage{verbatim}
\usepackage{amsmath}
\usepackage{amsfonts}
\usepackage{amssymb}
\usepackage[usenames]{color}
\usepackage{paralist}
\usepackage{ifthen}
\usepackage{tabularx}
\usepackage{multirow}
\usepackage{wrapfig}

\usepackage[dvipsnames]{xcolor}
\usepackage{colortbl}
\definecolor{mygray}{gray}{0.8}
\definecolor{mycol}{rgb}{1,0.9,0.9}
\definecolor{mcy}{rgb}{0.9,1,0.9}
\newtheorem{theorem}{Theorem}[section]

\newboolean{arxiv}
\setboolean{arxiv}{false}

\DeclareMathSymbol{\R}{\mathord}{AMSb}{"52}
\DeclareMathSymbol{\C}{\mathord}{AMSb}{"43}
\DeclareMathSymbol{\Z}{\mathord}{AMSb}{"5A}
\DeclareMathSymbol{\N}{\mathord}{AMSb}{"4E}
\DeclareMathSymbol{\K}{\mathord}{AMSb}{"4B}
\DeclareMathSymbol{\M}{\mathord}{AMSb}{"4D}
\DeclareMathSymbol{\Q}{\mathord}{AMSb}{"51}
\DeclareMathSymbol{\Lset}{\mathord}{AMSb}{"4C}

\DeclareMathOperator*{\argmax}{arg\,max}

\theorembodyfont{\rmfamily}
\theoremheaderfont{\bfseries\slshape}
\newtheorem{prop}{Proposition}
\newtheorem{corollary}{Corollary}
\newtheorem{lem}{Lemma}

\def\argmax{\operatornamewithlimits{arg\,max}}
\definecolor{webmag}{rgb}{0.5,0,0.5}
\definecolor{myblue}{rgb}{0,0,1}
\newcommand{\nop}[1]{}

\title{\huge Performance-Oriented Association in Large Cellular Networks \\with Technology Diversity}

\author{\IEEEauthorblockN{Abishek Sankararaman}
\IEEEauthorblockA{Dept. of ECE,
Univ. of Texas at Austin}
\and
\IEEEauthorblockN{Jeong-woo Cho}
\IEEEauthorblockA{School of ICT,
KTH, Sweden}
\and
\IEEEauthorblockN{Fran{\c c}ois Baccelli}
\IEEEauthorblockA{Dept. of ECE,
Univ. of Texas at Austin}
\and
}
\begin{document}
\maketitle

\newcommand{\expectation}{\textsf{E}}
\newcommand{\probability}{\textsf{P}}
\newcommand{\pdf}{\textsf{f}}
\newcommand{\slow}{\ell}
\newcommand{\nextline}{\mbox{}\\}
\newcommand{\ud}{\mathrm{d}}
\newcounter{tempcounter}
%\newcounter{acounter}s
\newcounter{acounter}

\begin{abstract}
The development of mobile virtual network operators,
where multiple wireless technologies (e.g. 3G and 4G) or operators
with non-overlapping bandwidths are pooled and shared
is expected to provide enhanced service with broader coverage,
without incurring additional infrastructure cost.
However, their emergence poses an unsolved question on how to
harness such a technology and bandwidth diversity.
This paper addresses one of the simplest questions
in this class, namely, the issue of associating each mobile
to one of those bandwidths. %These technologies can
%be those pooled by a MVNO or even those maintained by a single operator.
Intriguingly, this association issue  is intrinsically distinct from those in  traditional networks.
We first propose a generic stochastic geometry model lending itself to
analyzing a wide class of association policies
exploiting various information on the network topology, e.g. received pilot powers and fading values.
This model firstly paves the way for tailoring and designing an optimal association scheme to maximize \emph{any} performance metric of interest
(such as the probability of coverage) subject to the information known about the network. In this class of optimal association, we prove a result that the performance improves as the information known about the network increases.
%  which is gradually enhanced with respect to the amount of information.
Secondly, this model is used to quantify the performance of \emph{any} arbitrary association policy and not just the optimal association policy.
%Secondly, as a pragmatic epitome of our approach, we consider a limiting regime, \ie, high path loss case, which leads to a simplified
We propose a simple policy called the Max-Ratio which is not-parametric, i.e. it dispenses with the statistical knowledge of base station deployments commonly used in stochastic geometry models. We also prove that this simple policy is  optimal in a certain limiting regime of the wireless environment.
%We also  prove the asymptotical optimality of a pragmatic association scheme, called max-ratio, dispensing with statistical knowledge used in stochastic geometry models capturing base station deployments.
Our analytical results are combined with simulations to compare
these policies with basic schemes, which provide insights into (i) a practical compromise between performance gain and cost of estimating information and; (ii) the selection of association schemes under environments with different propagation models, i.e. path-loss exponents.

%In this paper, we study cell association in multi-technology networks, wherein a mobile-phone or  User Equipment (UE) has the choice of which  `technology' to associate with in addition to choosing the base-station from that technology. The base-stations (BS) across different technologies operate on orthogonal frequency bands and do not interfere with each other. For instance, one can think of the  technologies as different operators who operate on orthogonal frequency bands and have pooled together to create a common service for the users as in the Google Fi project. We present a  mathematical framework of such networks to  study various association policies including the `optimal' association. The optimal association policy is the one that maximizes a UE's reward among all policies that have the same data or `information' (to be made precise in the sequel) at the UE. We also propose the `Max-Ratio' policy, which is  more practical and easy to implement compared to the optimal policy. We then do performance analysis of the association policies we propose.
% We observe through simulations, the `diminishing returns' of performance of the optimal policy with increasing amount of data a UE posses about the network.

\end{abstract}

\section{Introduction}\label{sec:intro}

In traditional operated mobile networks, each user (mobile) is obliged to
subscribe to a particular operator and has access to the base stations
owned by the operator (or to Wi-Fi access points
administered by the operator). A new paradigm known as
{\itshape mobile virtual network operators} (MVNO) is currently reshaping
the wireless service industry.
The idea is to provide higher service quality and connectivity
by pooling and sharing the infrastructure of multiple wireless
networks.
A recent remarkable entrant such as Google is testing the
water in the US market under the name of ``Project Fi'', whose
main feature is improved coverage provided through outsourcing infrastructure from its partners, T-Mobile, Sprint and
their Wi-Fi networks. In the meantime, the European Commission
has been ruling favorably for MVNOs since 2006, so as to make
the European wireless market more competitive \cite{refECSpanish},
thereby facilitating investment in MVNOs in Europe.
These virtual operators can take advantage of the hitherto
impossibility to cherry-pick different network operators
which use separate bandwidths, and even different wireless
access technologies, for improvement of user experience.
It is reported \cite{refMcKinsey} that the market share of
these operators, especially in mature markets, ranged from 10\%
(UK and USA) to 40\% (Germany and Netherlands) as of 2014.
However, these unprecedented diversities in terms of bandwidths and
wireless technologies raise a challenging question on how to
harness them in large-scale wireless networks.

In the rest of the paper, we use the terminology ``technology diversity''
to refer to (i) several networks operated on orthogonal
bandwidths and (ii) different cellular technologies (e.g. 3G and 4G), both of which can be shared by MVNOs.

 \begin{figure}[hs!]
  \centering
  \centerline{\includegraphics[width=6.5cm]{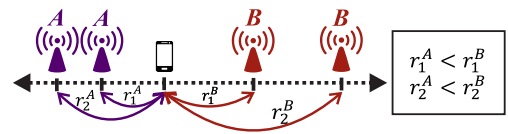}}
  \caption{Motivational example with two technologies: $r_1^A < r_1^B$ and $r_2^A < r_2^B$.} \label{fig:motivation}
\end{figure}

Notably, the {\itshape de facto} standard association policy in existing
wireless networks consists in
associating each user equipment (UE) with the nearest
base station (BS) or access point where one typical aim is
to maximize the likelihood of being covered or connected.
One of the main points of the present paper is that this
is no longer optimal in these emerging virtual networks,
as further discussed in Section \ref{sec:related}.
The subtle distinction arising from technology diversity
is illustrated in Fig. \ref{fig:motivation},
where $r_1^A$ and $r_2^A$ (respectively, $r_1^B$ and $r_2^B$)
denote the distances to the nearest and second-nearest BSs of technology $A$ (respectively, technology $B$)
from the UE located at the origin. Also, we assume that
$r_1^A < r_1^B$, i.e. the nearest BS of technology $A$
is the nearest to the UE, and there
are only four BSs as shown in Fig. \ref{fig:motivation},
which are identical except that they operate on different
technologies (i.e. non-overlapping bandwidths).
In the single technology case ($A=B$), the UE can simply
 associate with the BS at $r_1^A$. However, if $A \neq B $, e.g. the two
technologies operate on different bandwidths, the locations of
the strongest interferers, $r_2^A$ and $r_2^B$ (the second-nearest
BSs), may {\itshape overturn} the choice of technology $A$
when the strongest interferer of technology $B$ is much farther
from the UE than that of technology $A$, i.e. $r_2^B \gg r_2^A$,
thus boosting the signal-to-noise-ratio (SINR) of technology $B$.
In light of this example,
%and as we will see in more detail below,
optimal association in such networks requires sophisticated
policies adaptively exploiting available information.

 We can further generalize the above example and envisage a practical scenario where each UE can obtain the information about several received pilot signals of nearby BSs, as in 3G and 4G cellular networks, which can be translated into a vector of distances. In this paper, we are interested in investigating the following question.
 %Apart from the equivocal nature of the association policy epitomized in the above example, the performance metric (e.g. coverage probability) pursued by each MVNO may drastically differ with their service and pricing schemes, which raises the following question:

 \noindent Q: {\itshape How much performance gain is achievable { theoretically} by tailoring the association policy and how much of it can we achieve in practice by exploiting { available} information?}

\vspace{1mm}
\noindent{\bfseries Main Contributions}:
%such association policies without encapsulating these diversities present on a vast wireless network in a mathematically tractable manner, where each node or BS causes interference to the entire network.
To tackle this association problem, we propose a stochastic geometry
model of multi-technology wireless networks which partly builds upon
\cite{refAndrewTractable}. This leads to a generic analytical
framework lending itself to associating UEs to BSs
in such a way that various performance metrics are optimized in the
presence of the diversities alluded to above, and for
various degrees of available information at the UE.
In theoretical terms, the proposed framework
%large-scale wireless networks without resorting to exhaustive simulations and {\itshape designing} an association policy whose balance between the amount of available information and the performance index of interest is struck in a methodological fashion, underlined by the {\itshape
paves the way for {\itshape structural} results
on the partial ordering of optimal policy performance
(see Section \ref{sec:opt_association}) and a  methodology for quantifying the performance of various association policies in a mathematically tractable manner.
From the practical viewpoint, the results provide a mathematical edifice not only replacing exhaustive simulations but also usable for instance to analyze parsimonious association scheme,
such as the {\itshape max-ratio} algorithm defined in the paper.
We also prove {\itshape asymptotic optimality} of this pragmatic policy, which uses
only the distances to the nearest and second-nearest BSs.
%for each technology.
Remarkably, all association schemes
discussed in this work are underpinned by a user-centric approach
leveraging the information about the network that is
typically available at each UE in existing networks,
thereby dispensing with any need for centralized coordination.

%While the main idea of our work could be potentially extended to Wi-Fi networks, we are focused on cellular networks in the rest of the paper.

In the rest of the paper, after discussing the specificity of our problem with
respect to previous work, %in Section \ref{sec:related},
we describe
the notion of information exploited for the association in Sections \ref{sec:model} and
\ref{sec:opt_association} in order
to  characterize optimal
association policies which in turn ameliorate performance indices, which are %a parametric class of policies
founded upon an underlying stochastic model of BSs
and diverse types of information including fading values and distances
to the BSs across different technologies. After establishing the optimality of the  max-ratio algorithm under a limiting regime and deriving  a versatile formula for computation of resulting performance in Sections  \ref{sec:max_ratio} and \ref{sec:perf_analysis},
%We also propose other non-parametric yet pragmatic association policies
%in Section \ref{sec:asso_examples}, which are oblivious to the
%stochastic assumptions on the network.
we derive tractable
expressions for performance metrics of several association schemes and evaluate them
 in Sections \ref{sec:computational_ex} and \ref{sec:simulations}. 
%%%
%%% Conferece Version
%{\bfseries Note that due to space limitation, proofs of most results are deferred to the full version of the paper ...}.
%%% Full Version
\emph {The proofs of all the results are deferred to the Appendix}.

\subsection{Related Work}\label{sec:related}

The policy of associating each UE to the nearest BS or
the BS with the strongest received power has been taken
for granted in the vast literature on cell association.
This is for instance the case in the stochastic geometry model
of cellular networks \cite{refAndrewTractable}. The rationale is clear.
This leads to the highest connectivity for each UE to choose
the nearest BS unless it is possible to exploit the
time-varying fading information, which is often unavailable in practice.
Even with the recent emergence of heterogeneous wireless networks,
also called {\itshape HetNet}, the rule is still valid in terms of
coverage probability. That is, a UE is more likely to be covered if it
associates with a BS whose received long-term transmission power (called pilot power)
is the strongest. A stochastic geometry model to exploit this
heterogeneous transmission powers of BSs belonging
to multiple tiers in HetNets along with {\itshape fading}
information has been investigated in \cite{refKtierHetero}.

However, from the perspective of load balancing between cells,
the rule is invalid in general because each UE might be better
off with a lightly-loaded cell rather than heavily-loaded one
irrespective of the distances to them. In particular, in HetNet scenarios,
it is important to distribute UEs to macro-cells and micro-cells so
that they are equally loaded. The optimal association  in the HetNet
setting is inherently computationally infeasible, i.e. NP-Hard,
whereas the potential gains from load-aware association schemes
are much higher \cite{optimization_asso}. To tackle this problem,
a few approximate or heuristic algorithms were proposed based on
convex relaxations \cite{optimization_asso, opti_2} and
non-cooperative and evolutionary games \cite{rat_games, game_2}.
Most of these algorithms are iterative in nature, requiring many
rounds of messaging between UEs and BSs for
their convergence.
% to equilibrium. %Parallely, many `one-shot' policies were proposed which are based on the smart heuristic of `biasing' (for ex. \cite{offloading}). Under this heuristic, a UE associates to the BS from which it sees the maximum average signal power, where the power from a small cell BS is `biased' by a multiplicative factor. This is done to ensure that the small cells are not under-loaded. Such single shot association schemes, though sub-optimal are very easy to implement. Indeed the current 3GPP standards has a version of biasing for BS association in their standards \cite{3gpp_bias}. From a mathematical perspective however, there is no principled way to select the bias values and is set in practice through extensive numerical simulations.
%\\

% Biasing is association based on maximum received power among all BSs with the small cells' received power biased to increase the chance of associating with a small cell. Indeed, the current 3GPP standards implements a version of the single-shot association policy based on the biasing heuristic \cite{3gpp_bias}.

It must be  stressed that the multiple technology setting
studied here is a largely  unexplored territory where the \emph{ validity} of the standard rule to associate with the nearest BS is undermined, which is unprecedented in the literature
as exemplified in Fig. \ref{fig:motivation}. Lastly, while there have been considerable work adopting stochastic geometry models for analyzing \emph {given} algorithms in large wireless networks, our work is a radical turnaround in the way of harnessing the model: we investigate new opportunities to tailor and design such algorithms to optimize the performance.

\section{Stochastic  Network Model}
\label{sec:model}

In this paper, we  consider adapting association schemes to ameliorate any performance metric in a  downlink cellular network that is a function of the  SINR received at  a single typical UE.
%In the sequel, we argue why analysis from the perspective of a single typical user is representative of the performance of the down-link cellular system which has been standard in the literature \cite{Baccelli_Book,refKtierHetero}.
To this aim, we first describe a generic stochastic model of the network and define the general performance metric that is induced by an association policy of the UE of interest, which are assumed to be decoupled from those of other UEs.\footnote{Note that extending this framework and results therein to the case where the association policy of a user is affected by those of other users (e.g. load-balancing in HetNet) is mathematically far more challenging and thus is left to future work.}
Note that we retain our stochastic network model in the most generic form for easier mathematical manoeuvrability of key results in Section \ref{sec:opt_association}, which in fact holds for for a large class of point processes (PPs). For instance, the information structure $\mathcal{F}_I$ is simplified later in Section \ref{sec:perf_analysis}.

\subsection{Network Model}
\label{sec:net_model}

We consider  $T$ different technologies where $T$ is finite. The BS locations of technology $i \in [1,T]$ are assumed to be a realization of a homogeneous Poisson-Point Process (PPP) $\phi_i$ on $\mathbb{R}^2$  of intensity $\lambda_i$ independent of other PPPs. The typical user, from whose perspective we perform the analysis, is assumed to be located at the origin, without loss of generality. Denote by $r_{j}^{i} \in  \mathbb{R}_{+}$ the distance  to the $j$th closest point of $\phi_i$ to the origin, or equivalently the $j$th nearest BS, where ties are resolved arbitrarily. Hence $r^{i}_{1}$ denotes the distance to the closest point (BS) of $\phi_i$  from the origin.

%\subsection{Signal Model}
Each BS of technology $i$ transmits at a fixed power $P_i$. The received power at a UE from any BS is however affected by fading effects and signal attenuation captured in the propagation model, typically through the path-loss exponent. We assume independent fading, i.e. the collection of fading coefficients $H_j^i$, which denotes the corresponding value from the $j$th nearest BS in technology $i$ to the UE,  are jointly independent and identically distributed according to some distribution function. We model the propagation path loss  through a non-increasing function $l_i(\cdot): \mathbb{R}_{+} \rightarrow \mathbb{R}_{+}$, where $i \in \{1,2,...,T \}$, i.e. the propagation model for each technology is determined by a possibly different attenuation function. Hence, the signal power received at the typical UE  from the $j$th BS of technology $i$  is $P_i  H^i_j  l_i(r^{i}_{j})$. For mathematical brevity, we henceforth consider the point process $\phi_i$ of technology $i$ where each point is  {\itshape marked} with an independent mark denoting the fading coefficient between the point (BS) to the origin (UE). We can assume that all the random variables  belong to a single probability space denoted by $(\Omega, \mathcal{F}, \mathbb{P})$ \cite{refBaccelliStochasticV1}.

%In such stochastic geometric model of cellular networks as described above, it is sufficient to asses the performance of the down-link system from the perspective of a single typical user. The theory of Palm Calculus \cite{refBaccelliStochasticV1} formalizes the connection between the performance of a single typical user and the average performance seen by many users, if there were a point-process of users distributed in space independent of the BS locations. Hence, we assume without loss of generality that there is just a single  user in the network  at the origin.

%Without loss of generality, we conduct analysis on a typical UE located at the origin.

\subsection{Information at a UE}

Another point at issue in this paper is the tradeoff between the cost of ``information'' available at UE and the performance gain attained by the association policy making use of that information. For easier presentation of results, e.g. Theorem \ref{thm:opt_association},
the notion of information is  encapsulated in a sigma-field $\mathcal{F}_I$ which is a  sub-sigma algebra of the sigma-algebra $\mathcal{F}$ on which the marked point processes $\phi_i$ are defined. A sub-sigma algebra $\mathcal{F}^{'}$ of $\mathcal{F}$ is such that $\mathcal{F}^{'} \subseteq \mathcal{F}$. %We further assume that all additional random variables needed for association algorithm are $\mathcal{F}_{I}$ measurable. 
An example of information is $\mathcal{F}_{I} = \sigma \left( \cup_{i=1}^{T} \phi_i(B(0,w))\right)$, which corresponds to the sigma-field generated by the point process up to distance $w$ from the origin. In other words, the UE can estimate BS locations of different technologies $r_j^i$ such that  $r_j^i \leq w$. % from itself  (such local information was defined and considered in \cite{info_aloha} for example).
%However, for the time being we keep this notion of information as an abstract filtration $\mathcal{F}_{I}$ of $\mathcal{F}$ and work out a structural theorem (Theorem \ref{thm:opt_association}). We will in the sequel (Section \ref{sec:asso_examples}) consider some practical examples of the type of information (i.e. the filtration $\mathcal{F}_{I}$) that a typical UE can have and perform some computations.

%In this abstract setup however, we will always assume that the typical UE has knowledge of the densities of the different technologies and the fact that they are independent PPPs. In other words, we allow for the policy $\pi$ to depend on the densities $\lambda_{i}$ and the fact that the different generated form a PPP. In the sequel, we will  consider policies where the knowledge of the densities or even the statistical properties of $\phi_i$ are not available.

\begin{table}
\begin{center}
\begin{tabular}{|c|l|}
\hline
Notation & Brief Description \\
\hline
\hline
%$T$ & Number of generated \\
%\hline
$\phi_i$ & Point process corresponding to technology $i \in [1,T]$ \\
\hline
$\lambda_i$ & Intensity (density) of point process $\phi_i$\\
\hline
$p_i(\cdot)$ & Performance function when associated with technology $i$ \\
\hline
$\mathcal{F}_{I}$ & Information available at the typical UE \\
\hline
$j_i$ & $\arg \sup_{j \geq 1} \mathbb{E}[ p_{i}(\text{SINR}_{0}^{i,j}) \vert \mathcal{F}_{I}] $ \\
\hline
$i^*$ & The technology chosen by an association policy \\
\hline
$\mathcal{R}^{\pi}_{I}$ & Average performance of association policy $\pi$ with $\mathcal{F}_I$\\
\hline
$\mathcal{R}^{\pi^*}_{I}$ & Average performance of optimal policy $\pi^*$ with  $\mathcal{F}_I$ \\
\hline
\end{tabular}
\end{center}
\caption{Table of Notation}
\label{table:notation}
\end{table}

\subsection{Association Policies}
\label{sec:policy_definition}

An association policy governs the decisions on which   technology and BS  the typical user (who is located at the origin) should associate with. More formally, an association policy $\pi$ is a measurable mapping, i.e. $\pi : \Omega \rightarrow [1, T] \times \mathbb{N}$ which is $\mathcal{F}_{I}$ measurable. As stated before, we assume that all additional random variables needed by the policy $\pi$ are $\mathcal{F}_I$ measurable. The interpretation of the policy $\pi$ being $\mathcal{F}_I$ measurable is that a typical UE decides to choose a technology and a BS to associate with based only on the information obtainable in the network. It is important to note that while our discussion in this  paper mainly revolves around optimal policies denoted by $\pi^*$, our methodology for the performance evaluation in Section \ref{sec:perf_analysis} can be applied for any (suboptimal) policy.

%The sigma-algebra $\mathcal{F}_{I} \subseteq \mathcal{F}$ is a filtration of the probability space and denotes the ``information" about the network the typical user has before associating. In addition, we also assume that all additional random variables needed by the policy $\pi$ are $\mathcal{F}_I$ measurable. This notion of information at a UE is detailed more in the sequel along with some practical examples.

\subsection{Performance Metrics}

All performance metrics considered in this work are functions of    SINR (Signal to Interference plus Noise Ratio) received at the typical UE. The SINR of the signal received at the origin from the $j$th nearest BS of technology $i$ is:
\begin{equation}
\text{SINR}_{0}^{i,j} = \frac{P_i H^i_j l_i(r^i_j)}{  N_{0}^{i} + \sum_{k \in \mathbb{N} \setminus \{j\}} P_i H^i_k l_i(r^i_k)   },
\nonumber
\end{equation}
where $N_{0}^i$ is the thermal noise power which is a fixed constant for each technology $i \in \{1,2,... ,T \}$. In order to encompass a general set of most useful performance metrics in wireless networks, the performance of different association policies are evaluated through  non-decreasing functions of the SINR observed at the typical UE. Formally, let $p_i(\cdot):\mathbb{R}_{+} \rightarrow \mathbb{R}_{+}$ be a non-decreasing function for each $i \in \{1,2,...,T \}$ which represents the metric of interest if the typical UE associates with technology $i$. Since $\pi$ takes values in two coordinates $[1,T] \times \mathbb{N}$ (Section \ref{sec:policy_definition}), we divide them into separate coordinates which are denoted by $\pi(0) \in [1, T]$ and $\pi(1) \in \mathbb{N}$, respectively corresponding to the technology and BS chosen by the policy. Then the performance of the association policy $\pi$ when the information at the typical UE is quantified by $\mathcal{F}_I$  is then given by:
\begin{align}
\mathcal{R}^{\pi}_{I} = \mathbb{E}[p_{\pi(0)} (\text{SINR}_{0}^{\pi}) ].
\end{align}
The subscript $I$ refers to the fact that the information present at the typical UE is $\mathcal{F}_I$. The performance metric $\mathcal{R}^{\pi}_{I}$ is averaged over all realizations of the BS deployments, fading variables, and any additional random variables used in the  policy $\pi$.

Two  well-known examples of performance metrics used in practice are coverage probability and average achievable rate. Coverage probability corresponds to setting the function $p_i(x) = \mathbf{1}(x \geq \beta_i)$, which is the chance that the SINR observed at a UE from technology $i$ exceeds a threshold $\beta_i$. The other common performance metric of interest, average achievable rate, is defined as $p_i(x) = B_i \log_2 (1  + x)$, where the parameter $B_i$ is  the bandwidth of technology $i$. All results  on optimal association policy and performance evaluation are stated on the assumption of a general function $p_i(x)$. %We resort to the two functions only in Section \ref{sec:computational_ex}.

\section{Optimal Association Policy}
\label{sec:opt_association}

%\subsection{Information at a UE}

The optimal association policy denoted by $\pi^{*}$  is
\begin{align}
\pi^{*}_{I} = \arg \sup_{\pi} \mathcal{R}^{\pi}_{I},	
\label{eqn:opt_asso_defn}
\end{align}
where the supremum is over all $\mathcal{F}_{I}$ measurable policies.  From a practical point of view, the optimal association policy is the one that maximizes the performance of the typical UE among all policies having the same ``information". In this setup of optimal association, however, we always assume that the typical UE has knowledge of the densities $\lambda_i$ of the different technologies and the fact that they are independent PPPs although several fundamental results can be easily extended to more general point processes. %In the sequel, we will  consider the more practical (but sub-optimal) policies which do not assume any knowledge on the statistical properties of $\phi_i$ or even their densities.

%The information present at the typical UE about the different technologies, is mathematically quantified in-terms of a filtration of the sigma-algebra $\mathcal{F}$ on which the marked point processes $\phi_i$ are defined \cite{Baccelli_Book}. A filtration $\mathcal{F}^{'}$ of $\mathcal{F}$ is such that $\mathcal{F}^{'} \subseteq \mathcal{F}$ is a sub-sigma algebra of $\mathcal{F}$. We also assume that all additional random variables needed for association are $\mathcal{F}_{I}$ measurable.  For the time being we keep this  abstract and work out a structural theorem. We will in the sequel consider some practical examples of the type of information (i.e. the filtration $\mathcal{F}_{I}$) that a typical UE can have and perform some computations. In this abstract setup however, we will always assume that the typical UE has knowledge of the densities of the different technologies and the fact that they are independent PPPs. In other words, we allow for the policy $\pi$ to depend on the densities $\lambda_{i}$ and the fact that the different technologies form a PPP. In the sequel, we will  consider policies where the knowledge of the densities or even the statistical properties of $\phi_i$ are not available.

Since, we are interested in maximizing an increasing function of the SINR of the typical UE, the optimal association rule is clearly to pick the pair of technology and BS which yields the highest  performance conditional on $\mathcal{F}_I$.

\begin{prop}

The optimal association algorithm when the information at the typical UE is given by the filtration $\mathcal{F}_{I}$ is such that
\begin{align}
\begin{split}
\pi^{*}_{I}(0) &= \arg \max_{i \in [1,T]} \sup_{j \geq 1}\mathbb{E}[ p_{i}(\text{SINR}_{0}^{i,j}) \vert \mathcal{F}_{I}], \\
\pi^{*}_{I}(1) &= \arg \sup_{j \geq 1} \max_{i \in [1,T]} \mathbb{E}[  p_{i}(\text{SINR}_{0}^{i,j}) \vert \mathcal{F}_{I}],
\end{split}
\label{eqn:opt_association2}
\end{align}
where the UE must pick the technology $\pi^{*}_{I}(0)$ and the $\pi^{*}_{I}(1)$-th nearest BS to the origin in $\phi_{\pi^{*}_{I}(0)}$.
%In the above equations, $p_{c}^{(i),j}(\beta)$ is the indicator function that the SINR from the $j$th nearest BS to the origin in $\phi_i$ exceeds $\beta$. i.e.
%\begin{align}
%p_{c}^{(i),j}(\beta) = \mathbf{1}\left( \frac{P_i h_1 l(r_{j}^{i})}{N_0 + \sum_{k \geq 1, k \neq j} P_i h_k l(r_{k}^{i})} \geq \beta \right).
%\end{align}

\end{prop}
%where $SINR_i$ is the maximum
%\begin{align}
%SINR_i = \frac{h_1 (r_{i}^{1})^{-\alpha}}{N_0 + \sum_{j \geq 2} h_i (r_{i}^{j})^{-\alpha} }
%\end{align}
%\\

The performance of the optimal association is
\begin{align}
\mathcal{R}^{\pi^*}_{I} =
 \mathbb{E}[   \sup_{j \geq 1} \max_{i \in [1,T]} \mathbb{E}[   p_{i}(\text{SINR}_{0}^{i,j})  \vert \mathcal{F}_{I}]  ].
 \label{eqn:perf_metric}
\end{align}
Since  $[1,T]$ and $\mathbb{N}$ are countable sets, the order of the maxima in \eqref{eqn:perf_metric} does not matter. An important point to observe is that the optimal association given in \eqref{eqn:opt_association2} depends on the choice of the performance metric $\{p_i(\cdot)\}_{i=1}^{T}$. Hence, the optimal association rule would be potentially different if one was interested in maximizing coverage probability as opposed to maximizing rate-related metrics for instance.
\subsection{Ordering of the Performance of the Optimal Association}

In this sub-section, we prove an intuitive theorem (Theorem \ref{thm:opt_association}) stating that ``more" information leads to better performance. To this aim, we need the following simple lemma.

\begin{lem}
Let $Y$ be any $\mathcal{E}$-valued R.V.   and $g(\cdot) : C \times \mathcal{E} \rightarrow \mathbb{R}$. %Then
\begin{align*}
\mathbb{E}[ \sup_{x \in C} g(x, Y)] \geq \sup_{x \in C} \mathbb{E}[g(x,Y)].
\end{align*}
\label{lemma_jenson}
\end{lem}
%The proof is standard and deferred to the Appendix.
%\begin{IEEEproof}
%We have that
%\begin{align}
%\sup_{x \in C}g(x,Y) &\geq g(x,Y) \text{ } \forall x \in C, \nonumber
%\end{align}
%and hence
%\begin{align}
%\mathbb{E}[\sup_{x \in C}g(x,Y)] \geq \mathbb{E}[g(x,Y)] \text{ } \forall x \in C.
%\label{eqn:local_lemma_1}
%\end{align}
%Since Equation (\ref{eqn:local_lemma_1}) is valid for all $x \in C$, we can pick the supremum on the RHS, i.e.
%\begin{align}
%\mathbb{E}[\sup_{x \in C}g(x,Y)] \geq \sup_{x \in C} \mathbb{E}[g(x,Y)] .
%\end{align}
%%
%%
%%
%%Let $\epsilon > 0$ be arbitrary. Let $x^{*} \in C$ be such that
%%\begin{equation}
%%\mathbb{E}[g(x^{*},Y)] \geq (1 - \epsilon) \sup_{x \in C} \mathbb{E}[g(x,Y)].
%%\label{eqn:lemma_epsilon}
%%\end{equation}
%%We have for each $y$ in $\mathcal{E}$, $\sup_{x \in C}g(x,y) \geq g(x^{*},y)$. Therefore, we have that
%%\begin{align}
%%\mathbb{E}[\sup_{x \in C}g(x,Y)] \geq \mathbb{E}[g(x^{*},Y)]   \geq (1 - \epsilon) \sup_{x \in C} \mathbb{E}[g(x,Y)]
%%\end{align}
%\end{IEEEproof}

\begin{theorem}
	If $\mathcal{F}_{I_1} \subseteq \mathcal{F}_{I_2}$, then $\mathcal{R}^{\pi^*}_{I_1} \leq \mathcal{R}^{\pi^*}_{I_2}$ where the association rule is the optimal one given in \eqref{eqn:opt_association2}.
	%In words, this means that in a scenario where the typical UE has more information (with more quantified through the sigma-field inclusion), then the coverage probability obtained by optimal association is larger.
	\label{thm:opt_association}
\end{theorem}
%\begin{IEEEproof}
%	\begin{align}
%	\mathcal{R}^{\pi^*}_{I_2} &= \mathbb{E}[    \max_{i \in [1,T]} \sup_{j \geq 1} \mathbb{E}[ p_{i}(\text{SINR}_{0}^{i,j}) \vert \mathcal{F}_{I_2}]   ] \nonumber \\
%	& \stackrel{(a)}{=} \mathbb{E}[   \mathbb{E}[    \max_{i \in [1,T]} \sup_{j \geq 1} \mathbb{E}[p_{i}(\text{SINR}_{0}^{i,j}) \vert \mathcal{F}_{I_2}]   \vert \mathcal{F}_{I_1} ]] \nonumber \\
%	& \stackrel{(b)}{\geq} \mathbb{E}[   \max_{i \in [1,T]} \sup_{j \geq 1}   \mathbb{E}[    \mathbb{E}[ p_{i}(\text{SINR}_{0}^{i,j})  \vert \mathcal{F}_{I_2}]    \vert \mathcal{F}_{I_1}      ]] \nonumber \\
%	& \stackrel{(c)}{=} \mathbb{E}[  \max_{i \in [1,T]} \sup_{j \geq 1}  \mathbb{E} [  p_{i}(\text{SINR}_{0}^{i,j})  \vert \mathcal{F}_{I_1}]  ]  = \mathcal{R}^{\pi^*}_{I_1}   \nonumber
%	\end{align}
%	where $(a)$ follows from the tower property of expectation, $(b)$ follows from  Lemma \ref{lemma_jenson} and  $(c)$ follows from the tower property of expectation and the fact that $\mathcal{F}_{I_1} \subseteq \mathcal{F}_{I_2}$.
%\end{IEEEproof}

%This structural theorem implies that having more information (with ``more'' quantified via the sigma-field inclusion) at a UE leads to better performance if the association scheme is the optimal one as given in  \eqref{eqn:opt_association2}.
This theorem establishes a partial order on the performance of the optimal policy under different information scenarios at the UE for any performance functions $\{p_i(x)\}_{i=1}^{T}$.

\subsection{Optimal Association in the Absence of Fading Knowledge}

The following lemma is quite intuitive and affirms that the optimal strategy for a UE in the absence of fading knowledge is to associate to the nearest BS of the optimal technology. %We defer the proof to the Appendix due to space constraints.

\begin{lem}
If the information $\mathcal{F}_I$  at the typical UE does {\bfseries not} contain the fading random variables, then $j_i = \arg \sup_{j \geq 1} \mathbb{E}[ p_{i}(\text{SINR}_{0}^{i,j}) \vert \mathcal{F}_{I}] = 1$ and hence   $\pi^{*}_{I}(1)  = j_{\pi^{*}_{I}(0) } =1$. %In words, the optimal  policy in the absence of fading knowledge is to associate to the nearest BS of technology $\pi^{*}_{I}(0)$.
\label{lem:jstar}
\end{lem}

\subsection{Examples of Information}
\label{sec:asso_examples}

One common class of information is the ``{\itshape locally estimated information}'' which a UE may attain through measurements of (i) received long-term receive pilot signals, which can be easily converted into distances of BS, and (ii) instantaneous received signals, from which fading coefficients can be computed. For example,  the knowledge of the distances to BSs no farther than $w$ from the UE is quantified through the sigma-algebra $\mathcal{F}_w = \sigma \left( \cup_{i=1}^{T} \mathcal{F}^{i}_{w} \right)$, where $\mathcal{F}^{i}_{w} = \sigma \left( \phi_i(B(0,w))\right)$ is the sigma algebra generated by the  stochastic process $\phi(B(0,w))$.
Furthermore, in case the UE is capable of estimating fading information, one can opt for the sigma-field  generated by the marked stochastic process $\phi_i(B(0,w))$, denoted as $\mathcal{F}_{w}^{i,H}$, where each point (BS) is {\itshape marked} with a fading coefficient between the BS and the UE. Here the superscript $H$ refers to the sigma-field generated by the marked point-process.

 In existing networks, the most practical example is the knowledge of the nearest $L$ BSs of each technology, denoted by $\mathbf{r}^{L}_{i} = [ r^{1}_{i}, .., r^{L} ]$, i.e. the $L$-dimensional vector  of the distances. In terms of sigma-algebra, it can be defined as $\mathcal{F}_{L} = \sigma\left(  \cup_{i=1}^{T} \mathcal{F}^{i}_{L}\right)$, where $\mathcal{F}^{i}_{L} = \sigma \left( \phi_i (B(0,r^{L}_{i}))\right)$ is the knowledge of the $L$ nearest BS of each technology. One particularly intriguing  scenario is  complete  information about the BS deployments, i.e. $L=\infty$. Denote by $\mathcal{F}_{\infty}$ the sigma-field for this information scenario and  $\mathcal{R}^{\pi^*}_{\infty}$ as the performance obtained by the optimal policy knowing the entire network. Since $\mathcal{F}_{\infty}$ is the maximal element among all sub-sigma algebras of $\mathcal{F}$, it follows from Theorem \ref{thm:opt_association} that $\mathcal{R}^{\pi^*}_{\infty}$ is the the \emph{ upper bound} of all achievable performances. To strike a balance between the performance of interest and estimation cost at UE, each MVNO can evaluate  $\mathcal{R}^{\pi^*}_{L}$ to see how much the association policy with $L$ distances stack up against the upper bound $\mathcal{R}^{\pi^*}_{\infty}$.

 \section{Max-Ratio Association Policy}

\label{sec:max_ratio}

 While the parametric framework in Section \ref{sec:opt_association} paves the way for designing the association policy maximizing various metrics, the optimal schemes encapsulated in \eqref{eqn:opt_asso_defn} and \eqref{eqn:opt_association2} are amenable to tractable analysis only with the knowledge about the underlying PPPs $\phi_i$, i.e. their intensities $\lambda_i$. On the other hand, it is less conventional at the present time, if not unrealistic, to assume that the densities $\lambda_i$ are available at the UE in a real network. More importantly, in certain deployment scenarios, it is highly likely that the BS distribution follows a non-homogeneous point process with density (intensity) varying with the location over the network, thereby invalidating the homogeneous PPP assumption.

From the computational perspective, the optimal association can often demand substantial processing power of the UE particularly when the resulting association tailored for a specified performance metric is not simplified  into a tractable closed-form expression. In this light, it is desirable to have  policies that are completely oblivious to any statistical modeling assumption on the network, i.e. minimalistic policies exploiting universally available information such as distances to BSs, which can be computed from received pilot signal powers in 3G and 4G networks. To address these issues, we propose a \emph{max-ratio} association policy. This policy has access to  the ratio $r^{i}_2 / r^{i}_1$ information for each technology $i$, i.e. the information $\mathcal{F}_{I} = \sigma \left(  \cup_{i=1}^{T} r^i_2 / r^i_1 \right)$. The max-ratio association is formally described by
\begin{equation}
i^{*} = \max_{i \in [1,T]} r^{i}_2 / r^{i}_1, \text{  }  j^{*} =1.
%\label{eqn:max_ratio_association}
\nonumber
\end{equation}

% The examples of  policies we present can be broadly classified as being either parametric policies or non-parametric policies. Parametric policies are those  relying on statistical assumptions, namely of $\phi_i$'s being independent PPPs and their densities being known. For instance, the optimal association policy in Equation (\ref{eqn:opt_association2}) is an example of a parametric policy. Non parametric  policies on the other hand are those that do not assume any statistical knowledge on the point-process $\phi_i$ and are completely `data-dependent'. We propose non-parametric policies even though they are sub-optimal since they can be practically implemented with greater ease. The first two examples we  present belong to the class of parametric  association policies while the third algorithm is a non-parametric association policy. We finally conclude with a fourth example by putting all the examples (parametric and non-parametric) as an instance of a generalized association policy. The performance analysis of the generalized association policy which  relies on the parametric assumptions on the network given in Section \ref{sec:net_model}, is postponed to Section \ref{sec:perf_analysis}.

This ratio maximization implies that we  place a high priority on  a technology where simultaneously the distance to the nearest BS $r_1^i$ is  smaller and that to the second-nearest BS $r_2^i$ is larger than other technologies. Note also that the above expression can be easily rearranged into the ratio of the received pilot powers of the nearest and second-nearest BSs when the BS transmission powers within each technology is the same. We show in Theorem \ref{lem:ratio_assymptote_opt} that although this policy {\itshape per se} is a suboptimal heuristic, it is optimal (in the sense of \eqref{eqn:opt_association2})  under a certain limiting regime of the wireless environment.

\begin{theorem}
Let the noise powers $N^i_0 = 0$ for all technologies $i$ and the performance function for all technologies $p_i(\cdot)=p(\cdot)$ for all $i$. Consider the family of \textbf{power-law} path-loss functions $\{l^{(\alpha)}(\cdot)\}_{\alpha>2}$ where $l^{(\alpha)}(x) = x^{-\alpha}$. Let $k$ be \textbf{any} integer greater than or equal to $2$. If the information at the UE is the $k$-tuple of the nearest  distances of each technology $i$ i.e. $\mathcal{F}_{I} = \sigma( \cup_{i=1}^{T} (r_{1}^{i} \cdots , r_{k}^{i}))$, then
\begin{align}
\pi^{*}_{\alpha}(0)  \xrightarrow[]{\alpha \rightarrow \infty} \argmax_{i \in[1,T]} \frac{r_{2}^{i}}{r_{1}^{i}} \text{   a.s.},
\label{eqn:main_conv_state}
\end{align}
where $\pi^{*}_{\alpha} $ is the optimal association as stated in \eqref{eqn:opt_association2}. Recall $\pi^{*}_{\alpha}(1)  = 1$, $\forall \alpha$ from Lemma \ref{lem:jstar}.
\label{lem:ratio_assymptote_opt}
\end{theorem}

This theorem states that max-ratio association is optimal in cases where the signal is drastically attenuated (i.e. large path-loss exponents) with distance, e.g., {\itshape metropolitan} or {\itshape indoor} environments where the exponent reach values higher than 4, e.g. $\alpha \in [4, 7]$. It is noteworthy that $\alpha$ at higher frequencies as in LTE networks tends to be higher (See, e.g. \cite[Chapter 2.6]{Andreas_Book} and references therein). In addition, another remarkable implication of this theorem is that it suffices for the asymptotic optimality to exploit the \emph {reduced} information $r_1^i/ r_2^i$ per technology in lieu of the given original information, i.e. $r_1^i$ and $r_2^i$. Also, any supplementary information on distances (or received pilot powers) to the third-nearest or farther BSs is  \emph {superfluous} and does not influence the optimality of the association. In Sections \ref{sec:computational_ex}, we show that this association brings about surprisingly tractable expressions for key performance indices. %Furthermore, we compare the performance of max-ratio association at finite path-loss regime with that of optimal association through simulations in Section \ref{sec:simulations}.

\section{Framework for Performance Analysis}
\label{sec:perf_analysis}

In  Section \ref{sec:opt_association}, we compared the performance of the optimal association policy under different information scenarios by establishing a partial order on them without explicitly computing the performance $\mathcal{R}^{\pi}$. However, in order to quantify its impact on $\mathcal{R}^{\pi}$ without resorting to exhaustive simulations, one is also interested in its explicit expression for a given policy $\pi$, which may be an optimal policy as in Section \ref{sec:opt_association} or a suboptimal one as in Section \ref{sec:max_ratio}. We  demonstrate how to explicitly compute $\mathcal{R}^{\pi}$ in an {\it automatic} fashion (in Theorem \ref{thm:cover_prob}) for any arbitrary policy $\pi$ belonging to a large class of polices, called generalized association, which constitutes another part of our contribution. %This is a large class of policies (which contain for instance the max-ratio and the optimal association) that are practically relevant.

\subsection{Generalized Association}
\label{sec:gen_asso}

In the rest of the paper, we restrict our discussion to a class of association policies $\pi$ that are optimized over information with a special structure $\mathcal{F}_I$, incorporating what is conventionally available in  cellular networks. That is, in order to answer the question posed in \ref{sec:intro}, we  assume  that the form of information that a UE has about each technology $i$ is  a vector $\mathbf{r}_i \in \mathbb{R}^{L}$. For instance, if the mobile is informed of the smallest two distances of each technology and their instantaneous signal powers, then $\mathbf{r}_i$ is a 4-dimensional vector with $2$ dimensions representing the distances and the other $2$ dimensions corresponding to the instantaneous fading powers. That is, we adopt this reduced notation as a surrogate for the sigma-algebra notation in Section \ref{sec:opt_association} for simplicity of the exposition.  %From the practical perspective, it is easy for a UE to learn about the network with the information encoded as a vector rather than as an abstract sigma-algebra. Hence, this framework of representing information as a vector  encompasses a large class of practically useful  scenarios.
Formally, we assume that the association policy $\pi = \{\pi_i(\cdot)\}_{i=1}^{T}$, according to which a mobile chooses a technology to associate with is given by
\begin{equation}
i^{*} = \arg \max_{i \in [1,T]} \pi_i( \mathbf{r}_i,\lambda_i),
\label{eqn:association_rule}
\end{equation}
 where $j_i$, the index of BS of technology $i$ to which the UE associates conditioned on selection of $i$, i.e. $i = i^{*}$, and $\mathbf{r}_i$ is the $L$-dimensional vector of observation for technology $i$.

 It is noteworthy that when the technologies are operated on overlapping bandwidths,  the above form of association may be extended to a more general form $\pi_i( \{\mathbf{r}_i\}_{i=1}^{T},\{\lambda_i\}_{i=1}^{T})$, where each association policy utilizes not only the information regarding technology $i$ but also that about all other technologies. Envisioning this extension is easily justifiable because the {\itshape desirability} of technology $i$ (represented by $\pi_i(\cdot)$) is affected by the interference inflicted by other technologies. However, we leave it as future work and focus our discuss onto the restricted class of information $\mathcal{F}_I$ in \eqref{eqn:association_rule} which covers most interesting scenarios in case of non-overlapping frequency bandwidths.
 %but nonetheless cover most cases of interest. Indeed, a more general version of association would be $i^{*} =  \pi( \{\mathbf{r}_i\}_{i=1}^{T},\{\lambda_i\}_{i=1}^{T})$. However, Equation (\ref{eqn:association_rule})  models many policies of interest (like the Max-Ratio and the Optimal association policy) and we restrict our study to this class from henceforth. %We will denote by a generalized policy $\pi$ to be the collection $\{\pi_i\}_{i=1}^{T}$ of functions and $\mathcal{R}^{\pi}_I$ to denote the performance of the policy $\pi$ under information $\mathcal{F}_I$.

% Note that the association policy in itself can be either parametric or non-parametric. If for instance, \\  $\pi_i(\mathbf{r}_i, \lambda_i) = \sup_{j \geq 1} \mathbb{E}[p_i(\text{\text{SINR}}_{0}^{i,j}) \vert \mathbf{r}_i ]$ and $j_i = \arg \sup_{j \geq 1} \mathbb{E}[p_i(\text{\text{SINR}}_{0}^{i,j}) \vert \mathbf{r}_i ]$ with $\mathbf{r}_i$ being some local knowledge of the nearest $L$ BSs, then $\pi_i$ is the optimal association policy which is parametric. On the other hand, if the functions $\pi_i$ do not depend on $\lambda$, and if the information vector of each technology $\mathbf{r}_i$ is non-parametric (a scalar of distance to the nearest BS for example), then $\pi_i$ is a non-parametric policy. Hence, the  generalized association framework is broad enough to capture a lot of interesting association scenarios.
 %including the optimal association case. We will compute the performance analysis of this generalized association in the sequel.

\subsection{Performance Computation of the Generalized Association}

  For each technology $i$, we denote by $f_i(\mathbf{r}_i)$ the probability density function (pdf) of the information vector $\mathbf{r}_i$ of technology $i$. For instance, if $L=1$ and each mobile has knowledge about the location of the nearest base-station $r_1^i$, then it follows from the property of a PPP that $f_i(\mathbf{r}_i)=f_i(r_1^i)$ is the Rayleigh distribution with parameter $1 / \sqrt{2 \pi \lambda_i}$. As for the max-ratio policy, $f_i(\mathbf{r}_i)= f_i([r_1^i, r_2^i])$ becomes the distribution of the nearest and second-nearest BSs characterized by the underlying PPP of technology $i$. We also denote by $f_{i}^{*}(\mathbf{r})$  the pdf of the  vector $\mathbf{r}_{i}$ \textit{conditioned} on the event that technology $i$ is selected, i.e. $i^{*} = i$.

  Denote by $f_{\pi_i}(\cdot)$  the pdf of $\pi_i(\mathbf{r}_i,\lambda_i)$ and by $F_{\pi_i}(\cdot)$  the cumulative density function (cdf) of $\pi_i(\mathbf{r}_i,\lambda_i)$. To put it simply, $f_{\pi_i} (\cdot)$ is the cdf of a {\itshape function} $\pi_i (\cdot)$ of the given information $\mathbf{r}_i$ rather than that of $\mathbf{r}_i$ itself. For example, in case of the max-ratio policy, $f_{\pi_i}(\cdot)$ is the distribution of $r_2^i / r_1^i$. To prove the main theorem, we first need to delineate the interplay between the distribution of {\itshape optimal} technology $f_i^*(\cdot)$, its original distribution $f_i(\cdot)$, and the (cumulative) distribution of the association policy $F_{\pi_i}(\cdot)$. We have the following lemma from a direct application of Bayes' rule and independence of the  point processes $\phi_i$.

%We are interested however in computing the joint distribution of $i^{*}$ and $\mathbf{r}_{i^{*}}$.

\begin{lem}
The probability density function $f_{i}^{*}(\mathbf{r})$ is given by
\begin{equation}
 f^{* }_{i}(\mathbf{r})  = f_i(\mathbf{r}) \cdot \frac{1}{\mathbb{P}[ i^{*} = i ]} \cdot \prod_{j=1,j \neq i}^{T} F_{\pi_j}( \pi_i(\mathbf{r},\lambda_i)) .
 \label{eqn:defn_wistar}
\end{equation}
\label{lem:defn_wistar}
\end{lem}

%\begin{IEEEproof}
%
%From the definition of $f_{i}^{*}(\mathbf{r})$, we have,
%  \begin{align}
% f^{* }_{i}(\mathbf{r}) \mathbf{dr} &= \mathbb{P}[ r \in \mathbf{dr} \vert i = i^{*}] \nonumber\\
% &=  \frac{\mathbb{P}[ \{r \in  \mathbf{dr} \}\cap \{i = i^{*}]\}}{\mathbb{P}[  i = i^{*}]} \nonumber \\
% & \stackrel{(a)}{=}  \frac{\mathbb{P}[ \{r \in  \mathbf{dr} \}\cap_{j \neq i} \{ \pi_j(\mathbf{r}_j,\lambda_j) \leq \pi_i(\mathbf{r}, \lambda_j)  ]\}}{\mathbb{P}[  i = i^{*}]} \nonumber  \\
% & \stackrel{(b)}{=} f_i(\mathbf{r})\prod_{j=1,j \neq i}^{T} F_{\pi_j}( \pi_i(\mathbf{r},\lambda_i)) \frac{1}{p_{i}} \mathbf{dr},
% \end{align}
% where $p_{i}$ is the probability that $i = i^{*}$ and $\mathbf{dr}$ is an infinitesimal element of $\mathbb{R}^{L}$. Here $(a)$ follows from the definition of $i^{*}$ in equation (\ref{eqn:association_rule}) and $(b)$ follows from the independence of the different point process and hence independence of the observation vectors $\mathbf{r}_j$.
% %Independence of $y_i$ across $i$ yields that $p_{i} = \int_{y \in \mathbf{R}} q_i(y) \prod_{j=1,j\neq i}^{T} Q_j(y) dy$. Note that $\sum_{i=1}^{T}p_i = 1$. The above calculations give us that
%% \begin{align}
%% w^{* }_{i}(\mathbf{r})  = w_i(\mathbf{r})\prod_{j=1,j \neq i}^{T} Q_j( f(\mathbf{r},\lambda_i)) \frac{1}{p_{i}}
%% \label{eqn:defn_wistar}
%% \end{align}
% \end{IEEEproof}

%The proof is deferred to the Appendix.
The following theorem finally presents a direct method for computing the performance of any generalized association policy $\pi$. Recall that the performance of a policy $\pi$ is given by $\mathcal{R}^{\pi}_I = \mathbb{E}[p_{i^{*}}(\text{SINR}_{0}^{i^*,j_{i^*}})]$.
%Note that the performance of an association policy is averaged over all randomness in the system, i.e. the BS locations, fading powers and the observations from each of the technologies.

\begin{theorem}
The performance of the association algorithm $\pi$ under information $\mathcal{F}_I$ denoted by $\mathcal{R}^{\pi}_{I}$ is given by-
\begin{equation}
 \sum_{i=1}^{T} \int_{\mathbf{r} \in \mathbb{R}^{L}}\mathbb{E} [ p_i(\text{SINR}_{0}^{i,j_i})  \vert \mathbf{r} ]  f_i(\mathbf{r})\prod_{j=1,j \neq i}^{T} F_{\pi_j}( \pi_i(\mathbf{r},\lambda_i))d\mathbf{r},
\label{eqn:coverage_prob_thm}
\end{equation}
%where
%\begin{align}
%q_c(j_i; \mathbf{r},\lambda_i, P_i) = \mathbb{E} [ p_i(\text{SINR}_{0}^{i,j_i})  \vert \mathbf{r} ]
%\label{eqn:defn_pci}
%\end{align}
%is the conditional performance obtained by associating to technology $i$ conditioned on the observation.
where $\mathbb{E} [ p_i(\text{SINR}_{0}^{i,j_i})  \vert \mathbf{r} ]$ corresponds to the performance obtained by associating to technology $i$ \emph{conditioned} on the information about technology $i$ the UE has is the vector $\mathbf{r}$.
\label{thm:cover_prob}
\end{theorem}
%\begin{IEEEproof} The performance of a policy $\pi_i(\cdot)$ in \eqref{eqn:perf_metric} becomes:
%\begin{align}
%& \mathcal{R}^{\pi}_{I}  = \mathbb{E}[p_{i^{*}}(\text{SINR}_{0}^{i^*,j_{i^*}})]   \nonumber\\
%& =\sum_{i=1}^{T} \mathbb{P}[i = i^{*}] \mathbb{E} [p_i(\text{SINR}_{0}^{i,j_i})\vert  i = i^{*}] \nonumber \\
%& \stackrel{}{=} \sum_{i=1}^{T} \mathbb{P}[i = i^{*}] \mathbb{E}_{\mathbf{r}_i}[\mathbb{E} [p_i(\text{SINR}_{0}^{i,j_i})\vert \mathbf{r}_i, i = i^{*}]] \nonumber\\
%& \stackrel{(a)}{=} \sum_{i=1}^{T} \mathbb{P}[i = i^{*}]  \int_{\mathbf{r} \in \mathbb{R}^{L}} \mathbb{E} [p_i(\text{SINR}_{0}^{i,j_i})\vert \mathbf{r}, i = i^{*}]  f^{* }_{i}(\mathbf{r}) dr \nonumber \\
%& \stackrel{(b)}{=} \sum_{i=1}^{T} \mathbb{P}[i = i^{*}]  \int_{\mathbf{r} \in \mathbb{R}^{L}} \mathbb{E} [p_i(\text{SINR}_{0}^{i,j_i})\vert \mathbf{r}]  f^{* }_{i}(\mathbf{r}) dr \nonumber \\
%& \stackrel{(c)}{=} \sum_{i=1}^{T}   \int_{\mathbf{r} \in \mathbb{R}^{L}} \mathbb{E} [p_i(\text{SINR}_{0}^{i,j_i})\vert \mathbf{r}]  f_i(\mathbf{r})\prod_{j=1,j \neq i}^{T} F_{\pi_j}( \pi_i(\mathbf{r},\lambda_i)) dr. \nonumber
%\end{align}
% We use the definition of $f_{i}^{*}(\mathbf{r})$ to perform the averaging over $\mathbf{r}_i$ on the event $i=i^*$ in step $(a)$. Step $(b)$ follows from the independence of $\phi_i$ across $i$ and hence we can drop the conditioning on $i = i^*$. Step $(c)$ follows from Lemma \ref{lem:defn_wistar}.
%\end{IEEEproof}

This theorem states that we need only two expressions, {\bfseries information} distribution $f_i(\cdot)$ and {\bfseries policy} distribution $f_{\pi_i} (\cdot)$, in order to derive the performance metric. As exemplified earlier, while $f_i (\cdot)$ is usually a simplistic expression thanks to properties of PPP, mathematical  manipulability of $f_{\pi_i} (\cdot)$ highly relies on the complexity of the association policy.

% The coverage probability of the association rule can then be written as
% \begin{align}
% P_{c}^{L}(\beta) &=  \sum_{i=1}^{T}\mathbb{P}[ \{ i^{*} = i\} \cap \mathbf{1}(\text{SINR}_i \geq \beta) ] \\
% & = \sum_{i=1}^{T} p_i \mathbf{P}[  \mathbf{1}(\text{SINR}_i \geq \beta) \vert i^{*} = i ] \\
%&= \sum_{i = 1}^{T} p_{i} \int_{\mathbf{r} \in \mathbf{R}^L} p_c(\mathbf{r},\lambda_i). w_{i}^{*}(\mathbf{r}) dr \\
% & = \sum_{i=1}^{T}\int_{\mathbf{r} \in \mathbf{R}^L} p_c( \mathbf{r}, \lambda_i)w_i(\mathbf{r})\prod_{j=1,j \neq i}^{T} Q_j( f(\mathbf{r},\lambda_i)) dr
% \label{eqn:coverage_prob}
% \end{align}
%
% where $p_c(\mathbf{r},\lambda_i) = \sup_{j \geq 1} \mathbb{E}[p_{c}^{(i),j}(\beta) \vert \mathbf{r} ]$ is the coverage probability of a typical mobile at the origin connected to the nearest BS of a PPP of intensity $\lambda_i$ conditioned on the observation vector for this technology as given by $\mathbf{r}$.
% \\

 %Hence, if one can have a closed form expression for the CDF $F_{\pi_i}(y)$ for a given association rule, then we have an expression for the performance metric $\mathcal{R}^{f}$ as given in equation (\ref{eqn:coverage_prob_thm}). In the sequel, we outline certain examples where one can leverage Theorem \ref{thm:cover_prob} to obtain closed form expressions.

 %However, the CDF $F_{\pi_i}(y)$ cannot in general be obtained in closed form.
 %Note that Theorem (\ref{thm:cover_prob}) is general and can be used to evaluate the performance of \emph{any} association scheme that has access to a vector of information per technology.

 \section{Computational Examples}
 \label{sec:computational_ex}

In this section, we leverage the results in Section \ref{sec:perf_analysis} to derive several performance metrics in {\itshape selected} practical scenarios where the association policy  utilizes information $\mathcal{F}_I$ restricted to a vector of distances to BSs ($ \mathbf{r}_i$) and BS densities ($\lambda_i$) as shown in \eqref{eqn:association_rule}. Note however that one can directly compute the performance ($\mathcal{R}^{\pi}_{I}$ in Theorem \ref{thm:cover_prob}) with the probability density function of any association policy ($f_i(\mathbf{r})$) by exploiting Lemma \ref{lem:defn_wistar}. We show that the resulting performance expressions are mathematically tractable and lend themselves to quantifying the performance of large-scale wireless networks.

For the rest of this section, we consider two representative metrics: (i) {\itshape coverage probability}  $p_i(x) = \mathbf{1}(x \geq \beta_i)$ and (ii) {\itshape average achievable rate} where, to simplify the exposition, we assume the bandwidths of different technologies are the same, i.e. $p_i(x) = p(x) = \log_2(1 + x)$. However, Theorem \ref{thm:cover_prob} can be used to compute the performance of any arbitrary non-decreasing function $p_i(\cdot)$. We also assume  that the fading variable $H_j^i$ is exponential, i.e. Rayleigh fading, with mean $\mu^{-1}$ and the path-loss function $l_i(r) = r^{-\alpha}$ for $\alpha > 2$  for all $i \in [1,T]$. These assumptions have often been adopted for analysis of wireless systems \cite{Andreas_Book} and espoused in stochastic geometry models \cite{refBaccelliStochasticV1,Baccelli_Book}. %To further simplify the expressions, we assume in a few analytical results  that the thermal noise in each technology is $0$, i.e. $N_0^i = 0$.

  Let us  denote by $c_p(j;\mathbf{r},\lambda,P,\beta)$ the coverage probability of a UE at the origin served by the $j$th nearest BS to the origin where the BSs are spatially distributed as a PPP of intensity $\lambda$. Here each BS transmits at power $P$ and we are interested in the probabilistic event that the received \text{SINR} exceeds the threshold $\beta$. The vector $\mathbf{r}$ denotes the vector of distances to BSs, based on which the association decision will be made. More formally,
 \begin{align}
 c_p(j;\mathbf{r},\lambda,P,N_0, \beta) = \mathbb{E} \left[ \mathbf{1} \left( \frac{P H_j}{N_0 + \sum_{k \neq j} P H_k} \geq \beta \right)  \bigg| \mathbf{r}\right.]\label{eqn:coverage_original}
 \end{align}
 Likewise, we denote by $r(j;\mathbf{r},\lambda,P)$  the expected rate received by a typical UE at the origin when it is being served by the $j$th nearest BS to the origin where the BSs are distributed as a PPP of intensity $\lambda$ and transmitting at power level $P$:
 \begin{align}
 r(j;\mathbf{r},\lambda,P,N_0) &= \mathbb{E} \left[ \log_2 \left( 1 +\frac{P H_j}{N_0 + \sum_{k \neq j} P H_k} \right) \bigg| \mathbf{r} \right] \nonumber \\
 &= \int_{t \geq 0} c_p(j; \mathbf{r}, \lambda, P, N_0, 2^{t}-1)dt.
 \label{eqn:rate_cover_from_cover_prob}
 \end{align}
Therefore, once we derive an expression for the coverage probability, the average achievable rate expression follows immediately from the calculation of  the simple integral  in \eqref{eqn:rate_cover_from_cover_prob}. In the sequel, we first compute {\itshape technology-wise} expressions,  \eqref{eqn:coverage_original} and \eqref{eqn:rate_cover_from_cover_prob}, which are in turn  plugged as $p_i(\cdot)$ into Theorem \ref{thm:cover_prob}  to yield  coverage probability metric $\mathcal{R}^{cp}$ and average achievable rate metric $\mathcal{R}^{r}$, respectively.

\subsection{Optimal Association Policy}\label{sec:example_opt_policy}

Recall that in the absence of knowledge of fading information, the optimal association policy is to choose technology $i^*$ such that:
 \begin{align}
i^* &= \arg \max_{i \in [1,T]} c_p(1; r_i, \lambda_i, P_i,N_0^i, \beta_i), \label{eq:covprobselection}\\
i^* &= \arg \max_{i \in [1,T]} r(1; r_i, \lambda_i, N_0^i, P_i), \label{eq:avgrateselection}
 \end{align}
respectively for coverage probability and average achievable rate. Note also that it follows from Lemma \ref{lem:jstar} that it is unconditionally optimal to choose the nearest BS for each technology, i.e. $j^* = 1$. Thus our discussion in this section is focused on the choice of technology $i$.
 %We are now ready to investigate selected scenarios exemplifying practical applications of Theorem \ref{thm:cover_prob}.

In this example, we investigate two cases where the UE has knowledge of the distances to the nearest $r_1^i$ or up to the second-nearest BSs $[r_1^i, r_2^i]$ along with the densities of technologies $\lambda_i$ while being oblivious to the information about fading $H_j^i$. In comparison with the standard rule to associate with the nearest BS, this example demonstrates how our proposed framework can be used not only to design an optimal association algorithm maximizing a performance index but also to compute the resulting performance improvements arising from the additional knowledge of distances and densities. The following theorem delineates, among all technologies $i \in [1,T]$, which technology yields the best coverage probability metric.

 \begin{theorem}
If the UE has the knowledge about $r_1^i$, for all $i \in [1,T]$, the association rule \eqref{eq:covprobselection} with the following expression maximizes the coverage probability:
\begin{multline}
c_p(1;r_1,\lambda , P,  N_0, \beta) = {\rm e}^{ -\mu \beta N_0 r_1^{\alpha} P^{-1}} \\ \exp \left(- 2 \pi \lambda  \int_{u = r_1}^{\infty} \frac{1}{1 + \beta^{-1}\left(u/r_1 \right)^{\alpha}} u du \right).
\label{eqn:coverage_1_conditional}
\end{multline}
If the UE has the knowledge about $r_1^i$ and $r_2^i$, for all $i \in [1,T]$, the association rule \eqref{eq:covprobselection} with the following expression maximizes the coverage probability:
 \begin{multline}
 c_p \left(1; [r_1, r_2], \lambda, P, N_0, \beta \right) = {\rm e}^{ -\mu \beta N_0 r^{\alpha}_{1} P^{-1}} \frac{1}{1 + \beta\left(r_1/r_2  \right)^{\alpha}} \\ \exp \left(- 2 \pi \lambda  \int_{u = r_2}^{\infty} \frac{1}{1 + \beta^{-1}\left(u/r_1 \right)^{\alpha}} u du \right).
 \label{eqn:lemm_2nd_dist}
  \end{multline}
  \label{thm:coverage}
 \end{theorem}
% \begin{proof}
% This again follows from Equations $(16.5)$ and $(16.6)$ of \cite{Baccelli_Book}. It is presented in the Appendix later on for completeness.
% \end{proof}
%

To better understand the practical implications of \eqref{eqn:coverage_1_conditional}, we can consider the case where the thermal noise and threshold terms are identical, i.e. $N_0^i=N_0$ and $\beta_i=\beta$. Since both the first and second factors in the right-hand side of \eqref{eqn:coverage_1_conditional} are decreasing functions with respect to $r_1$, the above policy gives preference to smaller $r_1^i$ among all technologies $i \in [1,T]$, which is in line with our intuition.

However, for approximately similar values of $r_1^i$, it also reveals that the optimal policy tends to choose technology $i$ with lower density $\lambda_i$ because the right-hand side of \eqref{eqn:coverage_1_conditional} decreases with $\lambda$. The observation is in best agreement with our intuition again because technology $i$ with high density $\lambda_i$ implies that there are more interfering BSs on the average. On the other hand, the standard rule leads to higher chance of association with the technology with large $\lambda_i$ because the nearest BS is more likely to belong to the technologies consisting of higher number of BSs. Thus it can be deduced that in case of heterogeneous BS densities, the standard rule leads to very poor coverage performance because of its tendency to associate with the most \emph{ populous} technology, whereas the above equation reveals the optimality of associating with \emph{ sparsely populated} technology, which sheds light on the complex optimization to be carried out by MVNOs. Likewise, the optimal policy exploiting the additional information of $r_2^i$ exhibits similar tendencies in \eqref{eqn:lemm_2nd_dist} while it prefers technology $i$ with larger $r_2^i$, thus pushing the strongest interference signal as far as possible.

In order to compute the optimal performance metric $\mathcal{R}^{cp}$ resulting from the association rule maximizing the coverage probability, we first need to derive the probability distribution of $ c_p(1;r,\lambda_i, P_i, N_0, \beta)$, which is in turn plugged into  Theorem \ref{thm:cover_prob}. The CDF $F_{\pi_i}(y) = \mathbb{P}[ c_p(1;r,\lambda_i , P_i, N_0, \beta) \leq y]$ can be simplified into the following expression by using the fact that the nearest distance $r$ of BSs distributed as a PPP is  Rayleigh distributed with parameter $\frac{1}{\sqrt{2 \pi \lambda_i}}$.
\begin{lem}
The CDF $F_{\pi_i}(\cdot)$ is given by
\begin{align}
F_{\pi_i}(y) = e^{-\ln\left( \frac{1}{y}\right)\frac{1}{2 } \left(\int_{v = 1}^{\infty} \frac{1}{1 + \beta^{-1}\left(v \right)^{\alpha}} v dv \right)^{-1}}.
\label{eqn:c(y)}
\end{align}
\label{lem:opt_nearest}
 \end{lem}
Finally, plugging \eqref{eqn:c(y)} into   \eqref{eqn:coverage_prob_thm}, we get the following theorem on  the coverage probability maximized by the optimal association policy.
\begin{corollary}
The coverage probability resulting from the optimal association exploiting the knowledge of $r_1^i$ is
\begin{multline}
\mathcal{R}^{cp} = \sum_{i=1}^{T} \int_{\mathbf{r} \in \mathbf{R}} c_p(1;r,\lambda_i, P_i,N_0^i,\beta_i)  2 \pi \lambda_i r e^{-\pi \lambda_i r^2} \\ \prod_{j=1,j \neq i}^{T} \left( \frac{1}{c_p(1;r,\lambda_j,P_j,N_0^j,\beta_j)} \right)^{-\frac{1}{2 } \left(\int_{v = 1}^{\infty} \frac{1}{1 + (\beta_j)^{-1}\left(v \right)^{\alpha}} v dv \right)^{-1} } \nonumber.
\end{multline}
\end{corollary}

%If the performance metric of interest was the rate-coverage, then the optimal association algorithm would be
%\begin{align}
%i^*  = \arg \max_{i \in [1,T]} r_c(1; r_i, \lambda_i, P_i).
% \end{align}
%Since, the instantaneous fading is unknown, $j^*= 1$ from Lemma \ref{lem:jstar}. However, computing $F_{\pi_i}(y)$ in this case cannot be done exactly in closed form. This example illustrates that the computational framework of Theorem \ref{thm:cover_prob} can be used only   if one can caompute the CDF $F_{\pi_i}(\cdot)$ in closed form.

\subsection{Max-Ratio Association Policy}
\label{subsec:maxratio}

%While our parametric framework laid out in Section \ref{sec:opt_association} and Theorem \ref{thm:cover_prob} makes it possible to tailor the association policy to optimize any desired performance index, as shown in Section \ref{sec:example_opt_policy}, its PPP assumption on BS distribution necessitates intensity information $\lambda_i$.

%On the contrary, the max-ratio policy dispenses with density information, whose {\itshape asymptotic} optimality for large path-loss $\alpha$ has been proven in Theorem \ref{lem:ratio_assymptote_opt}.

%The Max-Ratio policy is a simple non-parametric policy that does not need the density information on the different BS.
Recall that in the absence of fading information, the Max-Ratio algorithm described in Section \ref{sec:max_ratio} is to choose technology such that $i^*= \max_{i \in [1,T]} {r^{i}_{2}}/{r^{i}_{1}}$ with the nearest BS in the chosen technology, i.e. $j^*= 1$. Although we saw in Section \ref{sec:example_opt_policy} that the density information play a crucial role in performing optimal association, we know from Theorem \ref{lem:ratio_assymptote_opt} that the simple non-parametric policy of Max-Ratio is optimal in the limit of large path-loss. In this section, we also show that this policy is tractable and yields expressions for key performance metrics (Corollary \ref{cor:max_ratio_coverage} and \ref{cor:max_ratio_rate}). The simplistic form of the policy distribution $F_{\pi_i}(\cdot)$ in the following lemma alludes to ensuing tractable results in this section.

%It is remarkable that the max-ratio algorithm in the context of optimal association laid out in Theorem \ref{lem:ratio_assymptote_opt} can be construed as maximizing the metric $P_i (r^{i}_1)^{-\alpha} / ( P_i (r^{i}_2)^{-\alpha} ) $, i.e. the ratio of the desired signal to the strongest interference signal. Apart from this practical interpretation of the max-ratio algorithm, its simple condition to determine the optimal technology expectably leads to a simplistic expression for performance metrics, as hinted in the following lemmas.
\begin{lem}
The law $F_{\pi_i}(\cdot)$ for the max-ratio algorithm is:
\begin{align}
F_{\pi_i}(x) = \mathbb{P} \left[ r^{i}_{2} / r^{i}_{1} \leq x \right] =   1 -  1/ x^2 .
\end{align}
\label{lem:distance_ratio}
\end{lem}

\begin{corollary}
The coverage probability performance $\mathcal{R}^{cp}$ of the max-ratio algorithm is given by
\begin{equation}
  2\sum_{i=1}^{T} \int_{t \geq 1} c_p\left(1; \frac{r^{i}_{2}}{r^{i}_{1}} = t, \lambda_i , P_i, N_0^i, \beta_i \right)  \frac{1}{t^3} \left( 1 - \frac{1}{t^2}\right)^{T-1}   dt,
\end{equation}
where
\begin{multline}
c_p\left(1; \frac{r^{i}_{2}}{r^{i}_{1}} = t, \lambda_i , P_i, N_0^i, \beta_i \right) = \\  \int_{u=0}^{\infty} c_p\left(1; \left[u ,ut \right], \lambda_i, P_i,N_0^i, \beta_i \right) 2(\pi \lambda_i)^2 u^3 t^4 e^{-\lambda \pi (ut)^2}  du,
\label{eqn:cp_defn_ratio}
\end{multline}
where  $c_p\left(1; \left[u ,ut\right], \lambda_i, P_i,N_0^i, \beta_i \right) $ is given in \eqref{eqn:lemm_2nd_dist}.
\label{cor:max_ratio_coverage}
\end{corollary}
Since the max-ratio does not optimize a particular performance metric but merely compares the ratio $r_2^i/r_1^i$, the average achievable rate expression can be obtained directly from the integral transform in \eqref{eqn:rate_cover_from_cover_prob}, which in turn is plugged into Theorem \ref{lem:ratio_assymptote_opt} to yield the following corollary.
\begin{corollary}
The average achievable rate  of the max-ratio algorithm is
\begin{multline}
\mathcal{R}^{r} =  2\sum_{i=1}^{T} \int_{v \geq 0}\int_{t \geq 1} c_p\left(1; \frac{r^{i}_{2}}{r^{i}_{1}} = t, \lambda_i , P_i, 2^v-1\right)  \\ \frac{1}{t^3} \left( 1 - \frac{1}{t^2}\right)^{T-1}   dt dv,
\end{multline}
where $c_p\left(1; \frac{r^{i}_{2}}{r^{i}_{1}} = t, \lambda_i , P_i, 2^v-1\right)   $  is given in \eqref{eqn:cp_defn_ratio}.
\label{cor:max_ratio_rate}
\end{corollary}

 To get more intuition about the formula, we present the following corollary.

\begin{theorem}
In the Interference-limited regime (i.e. $N_0^i = 0$ for all $i \in [1,T]$), if the path-loss function is given by $l_i(r) = r^{-\alpha}$ for some $\alpha > 2$ and all $i \in [1,T]$, the coverage probability and the average achievable rate of the max-ratio algorithm are respectively given by
\begin{equation}
\textstyle \mathcal{R}^{cp} = \left\{     \begin{array}{ll}
\displaystyle\sum_{i=1}^{T} \int_{x= 0}^{1} \textstyle \frac{ 2(T-1) \cdot x^3 (1-x^2)^{T-2}}{1 + \beta_i^{2/\alpha} \phi(\alpha,\beta_i,x)} dx , & T \geq 2 \\
\frac{1}{1 + \beta_1^{2/\alpha} \phi(\alpha,\beta_1,1)}, &  T=1
\end{array} \right. ,
\label{eq:max-ratio-no-noise}
\end{equation}
\begin{equation}
\textstyle \mathcal{R}^{r} = \left\{     \begin{array}{ll}
 \displaystyle \sum_{i=1}^{T} \int_{x= 0}^{1} \int_{t \geq 0} \textstyle \frac{ 2(T-1) \cdot x^3 (1-x^2)^{T-2}}{1 + (2^t-1)^{2/\alpha} \phi(\alpha,2^t-1,x)}  dt dx , &  T \geq 2 \\
\displaystyle \int_{t \geq 0} \textstyle \frac{1}{1 + (2^t-1)^{2/\alpha} \phi(\alpha,2^t-1,1)} dt, &  T=1,
\end{array} \right.
\label{eq:max-ratio-rate-no-noise}
\end{equation}
where the function $\phi$ is given by
\begin{align}
\phi( \alpha, y, x) = \int_{u \geq y^{-2/\alpha}} \frac{1}{1 + x^{-\alpha} u^{\alpha/2}} du. \nonumber
\end{align}
\label{thm:max_ratio}
\end{theorem}

%The proof of this is deferred to the Appendix.
Since the case $T=1$ in the above theorem corresponds to the standard rule to associate with the nearest BS in the presence of only one technology, Equations \eqref{eq:max-ratio-no-noise} and \eqref{eq:max-ratio-rate-no-noise} reduce to much simpler expressions compared to those in the literature, e.g. Sections III-D and IV-C in the work \cite{refAndrewTractable}. On the other hand, as $T$ becomes larger, inside integrand in \eqref{eq:max-ratio-no-noise}, the distribution $2T(T-1)x^3 (1-x^2)^{T-2}$ (additional $T$ cancels out the summation operation) is gradually skewed toward the origin $x=0$, around which $\phi(\alpha,\beta_i,x)$ approaches 0. It is easy to show that the coverage probability $\mathcal{R}^{cp}$ approaches one with higher \emph {technology diversity}, i.e. $T \to \infty$. Though it is not realistic to envision such a large number of technologies or operators, from which each UE can cherry-pick its optimal BS, this theorem demonstrates how much UEs potentially benefit from the multiple bandwidths or technologies pooled by MVNOs.

Contrary to the standard association which tends to pick more populous technologies (i.e. large $\lambda_i$), giving rise to higher number of interferers, the max-ratio policy counterbalances this pathological behavior by ensuring that the strongest interferer $r_2^i$ is located relatively further. At the same time, the overall performance of max-ratio algorithm critically relies on large path-loss constant $\alpha$, whereas with this {\itshape caveat}, Theorem \ref{lem:ratio_assymptote_opt} states that the algorithm is \emph {asymptotically optimal} as $\alpha\to\infty$ for any increasing performance function $p_i(\cdot)=p(\cdot)$ in the interference-limited regime.

\section{Simulations and Numerical Results}
\label{sec:simulations}
In this section, we provide more insights into our framework and results by performing simulations and noticing their trends. In performing the simulations, we take as performance metrics, the coverage probability with $p_i(x) = \mathbf{1}(x \geq \beta_i)$ and the average rate with $p_i(x) = \log_2(1+x)$.

\subsection{Diminishing Returns  with Increasing Information}

%We  observe diminishing returns of the performance of optimal association with increasing information in Figure \ref{fig:diminishing_1}. We see that (not surprisingly) it is not beneficial to invest resources by a UE to learn a lot about the network as the performance gains diminish rather rapidly with increasing network information. In fact, for typical values of the path-loss exponent, knowing just the top $2$ nearest BS per technology is almost as good as knowing the entire network topology.

\begin{figure}[t]
\centering
\includegraphics[width=6cm]{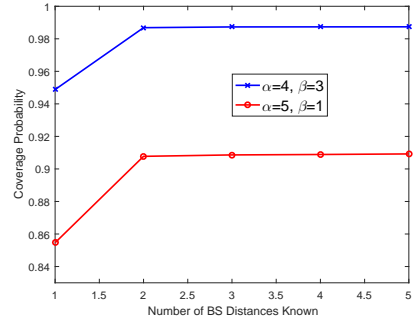}
\caption{The diminishing returns with increasing information with $T = 10$, the path-loss function $l_i(r) = r^{-\alpha}$, and the performance $p_i(x) = \mathbf{1}(x \geq \beta)$. }
\label{fig:diminishing_1}
\end{figure}

We first observe through simulations that the optimal association for any good class of performance metrics (made precise in the sequel) exhibits the law of diminishing returns. The term, diminishing returns, is used in the context where the additional gains or improvements in performance of optimal association reduces as the information increases. Fig. \ref{fig:diminishing_1} shows that coverage probability with the {\itshape optimal} association exploiting the knowledge of the nearest $k$ BS of each technology. As we move on the x-axis, we are increasing the information known at the UE and observe that the gains saturate drastically. Remarkably, beyond learning the $2$ nearest BSs per technology, there is no tangible improvement in the coverage probability. This implies that in practice, it is sufficient for each UE to learn the nearest two BSs per technology which will yield almost all the optimal performance possible with the full information about the topology.

We  present a simple argument  why one would expect to see diminishing returns for any performance metric. % and not  just the coverage probability. 
Assume we have some ``good" performance metric functions $\{p_i(\cdot)\}_{ i = 1}^{T}$, i.e. $\mathbb{E}[  p_{i}( \text{SINR}^{i,j}_{0})] $ is upper-bounded for all $i \in [1,T]$ and all $j \in \mathbb{N}$. %Most natural examples of performance metrics (e.g., average achievable rate and coverage probability) satisfy this criterion. 
Let $\{\mathcal{F}_{n} \}_{n \in \mathbb{N}}$ be a filtration of information $\mathcal{F}$ such that i.e. $\mathcal{F}_{n} \subseteq \mathcal{F}_{n+1} \subseteq \mathcal{F}$  for all $n$. Denote by $\mathcal{F}_{\infty}$ as the limit of $\mathcal{F}_n$, i.e. $\mathcal{F}_{\infty} = \cup_{n \geq 1} \mathcal{F}_n$ and let $\mathcal{R}^{\pi^*}_{n} = \mathbb{E} [  \sup_{j \geq 1} \max_{i \in [1,T]}   \mathbb{E} [  p_{i}( \text{SINR}^{i,j}_{0} ) \vert| \mathcal{F}_n  ]  ]$ be the performance of the optimal association policy under information $\mathcal{F}_n$. Theorem \ref{thm:opt_association} then gives that the sequence $\{ \mathcal{R}^{\pi^*}_{n} \}_{n \geq 1}$ is non-decreasing and $\mathcal{R}^{\pi^*}_{\infty} < \infty$. % since $\mathbb{E}[  p_{i}( \text{SINR}^{i,j}_{0})] < \infty$.
Any such  sequence of bounded and non-decreasing numbers contains a sub-sequence $\{ \mathcal{R}^{\pi^*}_{n_i}\}_i$ such that the gains $\Delta_{n_i} = R^{\pi^*}_{{n}_{i+1}} - R^{\pi^*}_{n_i}$ decreases with $i$. Therefore, the law of diminishing returns property holds.

\begin{figure*}[t!]
\begin{center}
\subfigure{\label{fig:ex1a}
        \includegraphics[width=5.8cm]{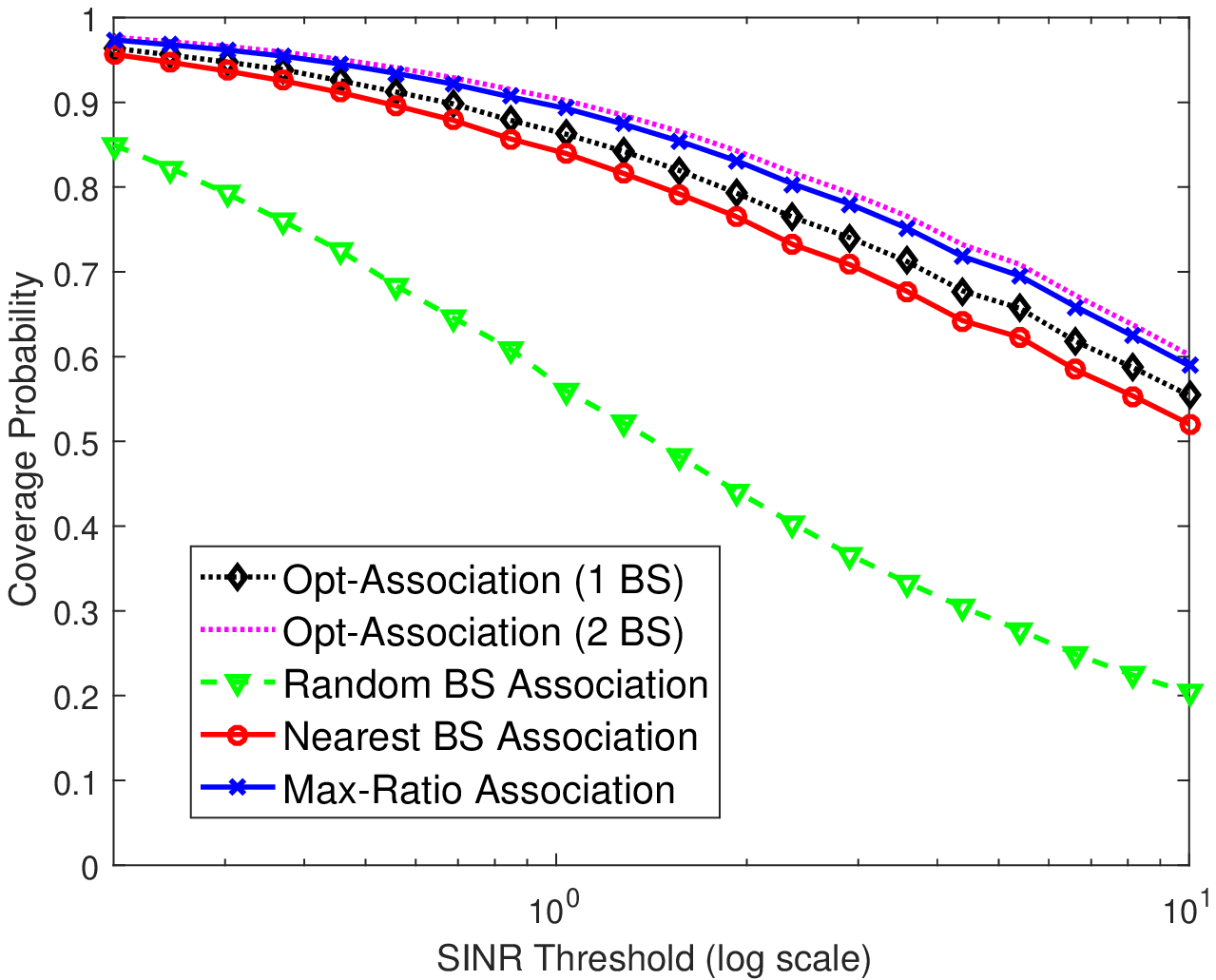}}
        \subfigure{\label{fig:ex1a}
        \includegraphics[width=5.8cm]{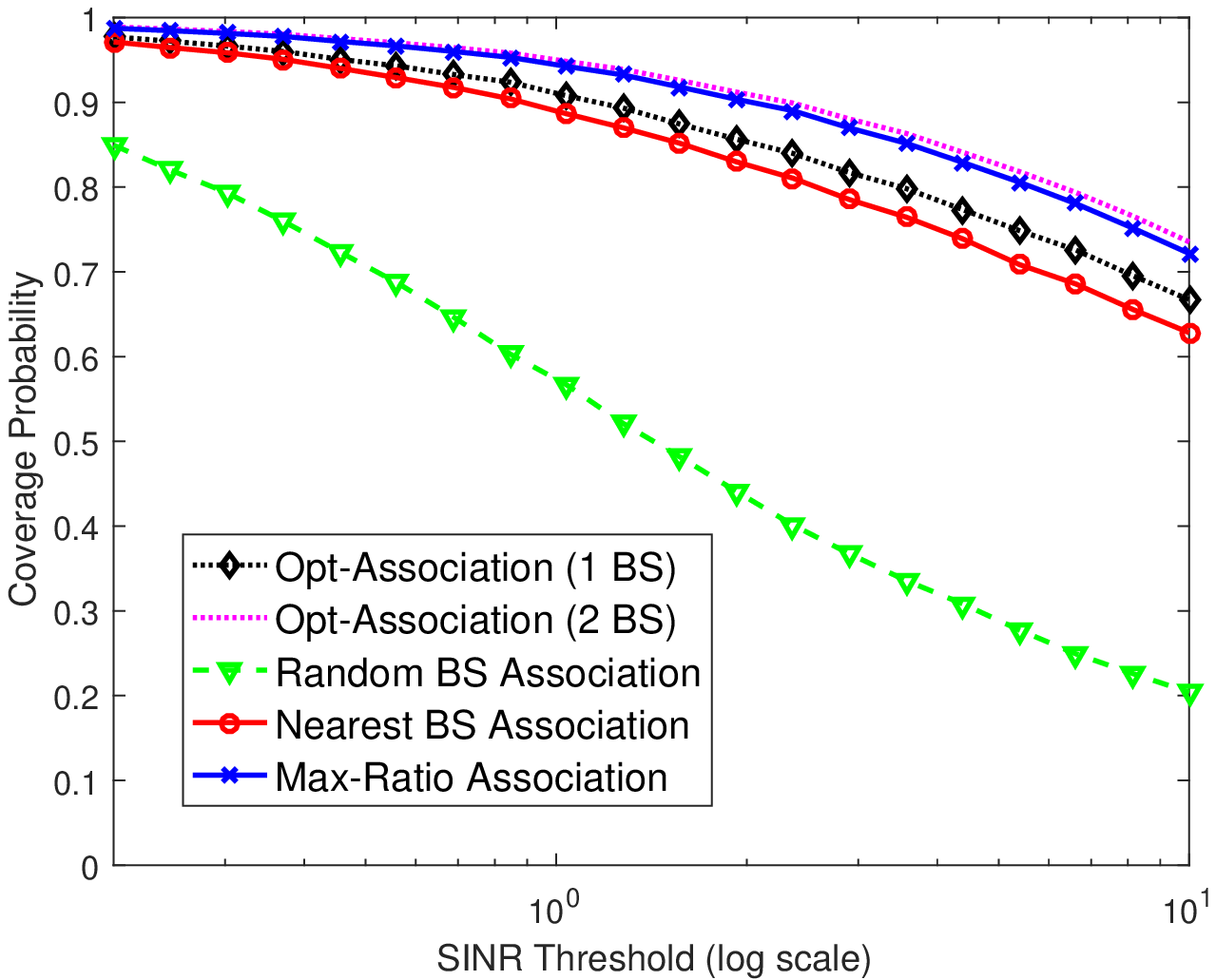}}
        \subfigure{\label{fig:ex1a}
        \includegraphics[width=5.8cm]{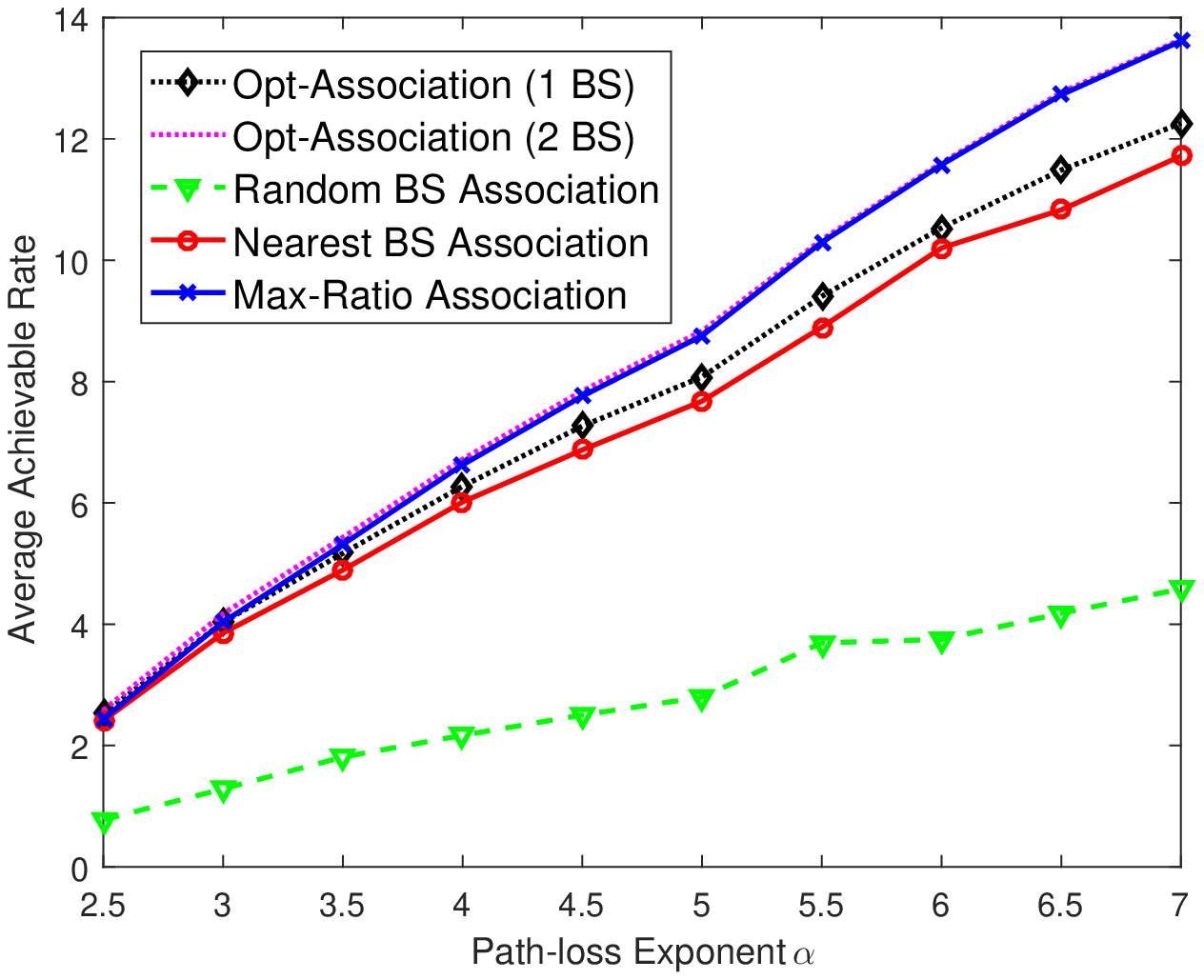}}
\caption{Comparison of various association schemes. The first two graphs on the left compare coverage probability where $l_i(r) = r^{-4}$ and $T=5$ for the first figure and $T=8$ for the second figure. The rightmost graph compares average achievable rate where $\alpha$ is varied on the x-axis and $l_i(r) = r^{-\alpha}$.}
\label{fig:comparision_1}
\end{center}
\end{figure*}

\subsection{Comparison of Schemes and Technology Diversity}

The first two graphs in Fig. \ref{fig:comparision_1} compare the coverage probability  of various association schemes with path-loss exponent $\alpha=4$ for different number of technologies, $T=5$ and $T=8$.  We observe in all graphs that the Max-Ratio association scheme  outperforms the optimal association policy under the case when only the nearest BS distances are known. More importantly, the Max-ratio association performs {\itshape almost as well} as the optimal association under the knowledge of nearest $2$ BSs per technology for this typical value of path-loss exponent, not to mention that it outperforms the nearest BS association significantly, particularly when the technology diversity is higher, i.e. $T=8$.

The rightmost graph in Fig. \ref{fig:comparision_1} depicts the average achievable rate for path-loss exponents $\alpha \in [2.5, 7]$, which empirically corroborates the statement of Theorem \ref{lem:ratio_assymptote_opt} that Max-Ratio is the optimal policy when nearest $k \geq 2$ BS per technology are known in the high path-loss regime. Remarkably,  Max-Ratio and the optimal association with two nearest BS distances performs almost equally  (indistinguishable in the graph) for $\alpha \geq 5$. That is, a simple non-parametric policy like the max-ratio performs  as well  as the optimal association policy in which the entire network topology is known (the best possible performance) even in the finite path-loss case. It is also noted that the random BS association, which is the only policy oblivious to technology diversity, results in poor performance in all cases. Thus it is  beneficial for MVNOs to leverage the technology diversity in any possible manner by all means.

%\begin{figure}
%\centering
%\includegraphics[scale=0.2]{avg_rate_fine.eps}
%\caption{A figure comparing the performance with the average rate $p_t(x) = \log_2(1+x)$ as the performance metric. In this plot, $T=10$ and  $l_t(r) = r^{-\alpha}$ where $\alpha$ is varied on the X-axis.}
%\label{fig:comparision_2}
%\end{figure}

%\subsection{Technology Diversity}

\begin{figure}
\centering
\includegraphics[width=6cm]{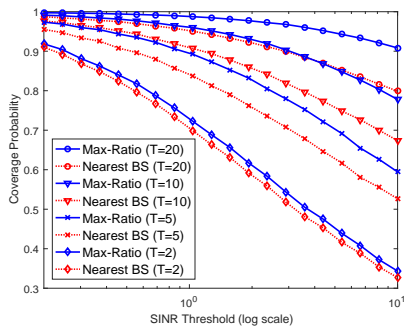}
\caption{Coverage probability increases with technology diversity $T$.}
\label{fig:coverage_inc_T}
\end{figure}

As shown in  Fig. \ref{fig:coverage_inc_T}, the coverage probability tends to one as $T$ goes to infinity, as discussed in Section \ref{subsec:maxratio}. The performance of Max-Ratio algorithm however reaches one quicker than nearest BS association. This shows that Max-Ratio exploits this diversity better than the conventional scheme to associate to the nearest BS.

%which gives Max-Ratio as a simple heuristic that exploits the technology better.

\section{Conclusions}

In this work, we explored the potential to boost the service performance of wireless networks without incurring additional infrastructure cost by capitalizing on a new form of diversity, which can be either several networks operated on orthogonal bandwidths or multiple wireless technologies pooled by some mobile virtual network operators.  We proposed a generic stochastic geometry model for designing association policies {\itshape proactively optimizing} desired performance metrics. We also showed that the most important metrics can in turn be evaluated  via a generic formula. Combined with another result characterizing the gradual increase of performance with respect to the amount of information, the framework provides a theoretical upper bound on the given metric, which can be used to determine the balance between the cost of estimating information at a mobile and the performance gain. Lastly, we devised a  pragmatic association scheme exploiting only two received pilot powers, whose asymptotic optimality is established under a limiting regime of high path-loss. As shown in the simulations, this scheme can serve as an alternative to the standard rules in urban or metropolitan environments with severe signal attenuation which better exploits the new form of diversity.

\section{Acknowledgements}
This work was supported by an award from the Simons Foundation (\# 197982) to The University of Texas at Austin.

{\footnotesize
\bibliographystyle{IEEEtran}
\bibliography{IEEEabrv,bib3}}

\appendix

\subsection{Proof of Lemma \ref{lemma_jenson}}
\begin{IEEEproof}
We have 
\begin{align}
\sup_{x \in C}g(x,Y) &\geq g(x,Y) \text{ } \forall x \in C, \nonumber
\end{align}
and hence
\begin{align}
\mathbb{E}[\sup_{x \in C}g(x,Y)] \geq \mathbb{E}[g(x,Y)] \text{ } \forall x \in C.
\label{eqn:local_lemma_1}
\end{align}
Since \eqref{eqn:local_lemma_1} is valid for all $x \in C$, we can pick the supremum on the RHS, i.e.
\begin{align}
\mathbb{E}[\sup_{x \in C}g(x,Y)] \geq \sup_{x \in C} \mathbb{E}[g(x,Y)] .
\end{align}
\end{IEEEproof}
\subsection{Proof of Theorem \ref{thm:opt_association}}
\begin{IEEEproof}
	\begin{align}
	\mathcal{R}^{\pi^*}_{I_2} &= \mathbb{E}[    \max_{i \in [1,T]} \sup_{j \geq 1} \mathbb{E}[ p_{i}(\text{SINR}_{0}^{i,j}) \vert \mathcal{F}_{I_2}]   ] \nonumber \\
	& \stackrel{(a)}{=} \mathbb{E}[   \mathbb{E}[    \max_{i \in [1,T]} \sup_{j \geq 1} \mathbb{E}[p_{i}(\text{SINR}_{0}^{i,j}) \vert \mathcal{F}_{I_2}]   \vert \mathcal{F}_{I_1} ]] \nonumber \\
	& \stackrel{(b)}{\geq} \mathbb{E}[   \max_{i \in [1,T]} \sup_{j \geq 1}   \mathbb{E}[    \mathbb{E}[ p_{i}(\text{SINR}_{0}^{i,j})  \vert \mathcal{F}_{I_2}]    \vert \mathcal{F}_{I_1}      ]] \nonumber \\
	& \stackrel{(c)}{=} \mathbb{E}[  \max_{i \in [1,T]} \sup_{j \geq 1}  \mathbb{E} [  p_{i}(\text{SINR}_{0}^{i,j})  \vert \mathcal{F}_{I_1}]  ]  = \mathcal{R}^{\pi^*}_{I_1}   \nonumber
	\end{align}
	where $(a)$ follows from the tower property of expectation, $(b)$ follows from  Lemma \ref{lemma_jenson} and  $(c)$ follows from the tower property of expectation and the fact that $\mathcal{F}_{I_1} \subseteq \mathcal{F}_{I_2}$.
\end{IEEEproof}
%
%
%
%Let $\epsilon > 0$ be arbitrary. Let $x^{*} \in C$ be such that
%\begin{equation}
%\mathbb{E}[g(x^{*},Y)] \geq (1 - \epsilon) \sup_{x \in C} \mathbb{E}[g(x,Y)].
%\label{eqn:lemma_epsilon}
%\end{equation}
%We have for each $y$ in $\mathcal{E}$, $\sup_{x \in C}g(x,y) \geq g(x^{*},y)$. Therefore, we have that
%\begin{align}
%\mathbb{E}[\sup_{x \in C}g(x,Y)] \geq \mathbb{E}[g(x^{*},Y)]   \geq (1 - \epsilon) \sup_{x \in C} \mathbb{E}[g(x,Y)]
%\end{align}

\subsection{Proof of Lemma \ref{lem:jstar}}

\begin{IEEEproof}

From the properties of PPP, we know that almost-surely, the distances $\{r^{i}_{k}\}_{k \geq 1}$ are distinct i.e.  satisfy $r^{i}_{k} > r^{i}_{k+1}$. Denoting $S_k = P_i l_i(r^{i}_{k})$ for each $k \in \mathbb{N}$ (instead of representing it as $S^{i}_{k}$, we drop the $i$ in this proof for simplicity), we can write   \eqref{eqn:opt_association2} as
\begin{align}
j_i  &= \arg \sup_{j \geq 1} \mathbb{E} \left[ p_i \left( \frac{S_j H_j}{N_{0}^{i} + \sum_{d \neq j} S_d H_d} \right)  \bigg| \mathcal{F}_I\right] \nonumber\\
& \stackrel{(a)}{=} \arg \sup_{j \geq 1} \mathbb{E} \left[ \frac{S_j H_j}{N_{0}^{i} + \sum_{i \neq j} S_i H_i} \bigg| \mathcal{F}_I\right], \nonumber \\
\end{align}
where $(a)$ follows from the fact that the function $p_i(\cdot)$ is non-decreasing.
\begin{align}
&\sup_{j \geq 1} \mathbb{E} \left[ \frac{S_j H_j}{N_{0}^{i} + \sum_{d \neq j} S_d H_d} \bigg| \mathcal{F}_I\right] \nonumber \\ &=  \sup_{j \geq 1} \mathbb{E} \left[ \mathbb{E} \left[\frac{S_j H_j}{N_{0}^{i} + \sum_{d \neq j} S_d H_d} \bigg| \sigma(\mathcal{F}_I \cup \phi_i) \right] \bigg| \mathcal{F}_I \right] \nonumber \\
& \leq \mathbb{E} \left[ \sup_{j \geq 1} \mathbb{E} \left[\frac{S_j H_j}{N_{0}^{i} + \sum_{d \neq j} S_d H_d} \bigg| \sigma(\mathcal{F}_I \cup \phi_i) \right] \bigg| \mathcal{F}_I\right],
\label{eqn:proof_lemma_1}
\end{align}
where the inequality follows from Lemma \ref{lemma_jenson}.  Since conditioned on $\phi_i$, we have $S_k$  deterministic and $H_k$  conditionally i.i.d. given $\phi_i$ and  independent of $\mathcal{F}_I$, we have 
\begin{align}
&\sup_{j \geq 1} \mathbb{E} \left[\frac{S_j H_j}{N_{0}^{i} + \sum_{d \neq j} S_d H_d} \bigg| \sigma(\mathcal{F}_I \cup \phi_i)\right] \nonumber \\ &= \sup_{j \geq 1} \mathbb{E} \left[{S_j H_j} \vert \sigma(\mathcal{F}_I \cup \phi_i)\right] \mathbb{E} \left[ \frac{1}{N_{0}^{i} + \sum_{d \neq j} S_d H_d} \bigg| \sigma(\mathcal{F}_I \cup \phi_i) \right] \nonumber \\
&= \sup_{j \geq 1} \mathbb{E} \left[{S_j H_j} \vert \phi_i\right] \mathbb{E} \left[ \frac{1}{N_{0}^{i} + \sum_{d \neq j} S_d H_d} \bigg| \phi_i \right].
\label{eqn:supremum_2}
\end{align}
Thus $j=1$ achieves the supremum in \eqref{eqn:supremum_2} since $S_k > S_{k+1}$ and is deterministic given $\phi_i$. Combining this fact with \eqref{eqn:proof_lemma_1}, we have
\begin{align}
\sup_{j \geq 1} \mathbb{E} \left[ \frac{S_j H_j}{N_{0}^{i} + \sum_{d \neq j} S_d H_d} \bigg| \mathcal{F}_I \right] \leq \mathbb{E} \left[ \frac{S_1 H_1}{N_{0}^{i} + \sum_{d \geq 2} S_d H_d} \bigg| \mathcal{F}_I\right],
\end{align}
which yields that $j_i = 1$. 
\end{IEEEproof}

\subsection{Proof of Theorem \ref{lem:ratio_assymptote_opt}}

\begin{IEEEproof}

For ease of notation, we denote the path-loss function as simply $l(\cdot)$ instead of $l^{(\alpha)}(\cdot)$, i.e. implicitly assume the dependence on $\alpha$ as  $l(x) = x^{-\alpha}$. For each fixed $\alpha$, we have from \eqref{eqn:opt_asso_defn}
\begin{align}
&\pi^{*}_{\alpha}(0) \nonumber\\  &= \argmax_{i \in[1,T]} \mathbb{E} \left[  p \left( \frac{P_i l(r^{i}_1) H^i_1}{  \sum_{j \geq 2} P_il(r^i_j)H^i_j } \right)  \bigg| (r^l_1,\cdots , r^l_k)_{l=1}^{T} \right] \nonumber \\
&= \argmax_{i \in[1,T]} \mathbb{E} \left[  \frac{P_i l(r^{i}_1) H^i_1}{  \sum_{j \geq 2} P_il(r^i_j)H^i_j }  \bigg|  (r^l_1,\cdots , r^l_k)_{l=1}^{T} \right] \nonumber \\
&=  \argmax_{i \in[1,T]} \mathbb{E} \left[   \frac{1}{   \sum_{j \geq 2}   \frac{l(r^i_j)H^i_j}{l(r^i_1)H^i_1}         }      \bigg| (r^i_1, \cdots , r^i_k) \right]. \label{eqn:lem_ratio_condexp}
\end{align}
We now argue that for each technology $i$, the conditional expectation in \eqref{eqn:lem_ratio_condexp} can be written as $A\frac{l(r^i_1)}{l(r^i_2)} - e^{(\alpha)}_{i}$ such that $e^{(\alpha)}_{i} \xrightarrow[]{\alpha \rightarrow \infty} 0$ almost surely and $A$ is a positive constant independent of $i$ and $\alpha$. If we show this, then the lemma can be proved as follows:
\begin{align}
\pi^{*}_{\alpha}(0)  &= \argmax_{i \in [1,T]} A\frac{l(r^i_1)}{l(r^i_2)} - e^{(\alpha)}_{i} \label{eqn:lem_ratio_wanted} \\
& \stackrel{(a)}{=} \argmax_{i \in [1,T]} A\frac{r^i_2}{r^i_1} - e^{(\alpha)}_{i} \nonumber \\
&\xrightarrow[a.s.]{\alpha \rightarrow \infty} \argmax_{i \in [1,T]} A\frac{r^i_2}{r^i_1}  = \argmax_{i \in [1,T]} \frac{r^i_2}{r^i_1}, \label{eqn:lem_ratio_lim2}
\end{align}
where step $(a)$ follows from the fact that $\frac{l(a)}{l(b)} = l(a/b)$ and the fact that $l(\cdot)$ is non-increasing. Since $ e^{(\alpha)}_{i}$ converges to $0$ almost-surely $\forall i \in [1,T]$, a finite set, we have uniform convergence i.e. $\sup_{i \in [1,T]}  e^{(\alpha)}_{i} \xrightarrow[]{\alpha \rightarrow \infty} 0$ almost-surely which gives \eqref{eqn:lem_ratio_lim2}.
 In the rest of the proof, we show that \eqref{eqn:lem_ratio_condexp} can be written as $A\frac{l(r^i_1)}{l(r^i_2)} - e^{(\alpha)}_{i}$.
 \\
 % such that $e^{(\alpha)}_{i} \xrightarrow[]{\alpha \rightarrow \infty} 0$ almost surely and $A$ is a positive constant not depending on $i$ and $\alpha$.

%\mathbb{E} \left[ \frac{l(r^i_1) H^i_1}{l(r^i_2) H^i_2} \bigg| (r^i_1, \cdots , r^i_k)  \right] - \\&  \frac{l(r^i_1) }{l(r^i_2) } \mathbb{E} \left[   \frac{H^i_1}{H^i_2} \left(\frac{Q^{(\alpha)}_{i}}{1 + Q^{(\alpha)}_{i}} \right) \bigg| (r^i_1, \cdots , r^i_k) \right], \nonumber \\
Expanding on the conditional expectation in \eqref{eqn:lem_ratio_condexp} using simple algebra to factor out the leading term, we get
\begin{multline}
\mathbb{E} \left[   \frac{1}{   \sum_{j \geq 2}   \frac{l(r^i_j)H^i_j}{l(r^i_1)H^i_1}         }      \bigg| (r^i_1,.., r^i_k)  \right]   \stackrel{(b)}{=}  \frac{l(r^i_1) }{l(r^i_2) }  \mathbb{E} \left[ \frac{H^i_1}{H^i_2} \bigg| (r^i_1,.., r^i_k) \right] -  \\  \frac{l(r^i_1) }{l(r^i_2) } \mathbb{E} \left[   \frac{H^i_1}{H^i_2} \left(\frac{Q^{(\alpha)}_{i}}{1 + Q^{(\alpha)}_{i}} \right) \bigg| (r^i_1, \cdots , r^i_k) \right],
\label{eqn:lem_ratio_conv_inter}
\end{multline}
where $Q^{(\alpha)}_{i} = \sum_{j \geq 3}\frac{l(r^i_j) H^i_j}{l(r^i_2)H^i_2}$. Step $(b)$ follows from the fact that $H^i_j$ are i.i.d. random-variables. Indeed \eqref{eqn:lem_ratio_conv_inter} resembles \eqref{eqn:lem_ratio_wanted} with the constant $A = \mathbb{E} \left[   \frac{H^i_2}{H^i_1} \right]$ (which is independent of $i$ and $\alpha$). It thus remains to prove that the second term (which is the error $e^{(\alpha)}_{i}$) in \eqref{eqn:lem_ratio_conv_inter} goes to $0$ almost surely as $\alpha$ goes to infinity.
\\

 From Campbell's Theorem, we know that
\begin{multline}
\mathbb{E}[ Q^{(\alpha)}_{i} \vert (r^i_1, \cdots , r^i_k) , H^i_2]  = \frac{\mathbb{E}[H] \sum_{z=3}^{k} l(r^i_z) }{H^i_2l(r^i_2)} + \\  \frac{\mathbb{E}[H] }{H^i_2l(r^i_2)} 2 \pi \lambda_i \int_{ u \geq r^i_k} l(u) udu.
\label{eqn:lem_ratio_camp1}
\end{multline}
with the notation that $\sum_{z=a}^{b} \cdot = 0$ if $b < a$. Furthermore,  since $l(x) = x^{-\alpha}$, we have $\frac{1}{l(\epsilon)}\int_{u \geq \epsilon}l(u)u du$ goes to $0$ as $\alpha$ goes to $\infty$ for every $\epsilon >0$. Thus, we have from \eqref{eqn:lem_ratio_camp1} and the fact that for a homogeneous PPP of positive intensity $\lambda_i$, $r^i_j > r^i_{j+1}$ a.s. $\forall j \in \mathbb{N}$, we get
  \begin{align}
  \lim_{\alpha \rightarrow \infty} \mathbb{E}[ Q^{(\alpha)}_{i} \vert (r^i_1, \cdots , r^i_k), H^i_2] =0 \text{ a.s.}
  \label{eqn:lem_ratio_camp}
  \end{align}

  Note that we needed to invoke Campbell's theorem, since we need to conclude about a sum of infinite random variables involved in the definition of $Q^{(\alpha)}$. Thus,
\begin{align}
e^{(\alpha)}_{i} &= \frac{l(r^i_1) }{l(r^i_2) } \mathbb{E} \left[   \frac{H^i_1}{H^i_2} \left(\frac{Q^{(\alpha)}_{i}}{1 + Q^{(\alpha)}_{i}} \right) \bigg| (r^i_1, \cdots , r^i_k) \right] \nonumber \\
&\leq \frac{l(r^i_1) }{l(r^i_2) } \mathbb{E}\left[ \frac{H^i_1}{H^i_2} \mathbb{E}[ Q^{(\alpha)}_{i} \vert (r^i_1, \cdots , r^i_k), H^i_1,H^i_2]  \bigg| (r^i_1, \cdots , r^i_k) \right ] \nonumber \\
& \stackrel{(c)}{} \xrightarrow[]{\alpha \rightarrow \infty} 0 \text{ a.s. } \label{eqn:lem_ratio_convergence2}
\end{align}
where step $(c)$ follows from \eqref{eqn:lem_ratio_camp} (through Dominated Convergence) and the fact that $H^i_1$ is a finite mean  random variable independent of everything else. Since $e^{(\alpha)}_{i}$ is positive, inequality \eqref{eqn:lem_ratio_convergence2} yields that $e^{(\alpha)}_{i} \rightarrow 0$ a.s.
\end{IEEEproof}

\subsection{Proof of Lemma \ref{lem:defn_wistar}}

\begin{IEEEproof}
From the definition of $f_{i}^{*}(\mathbf{r})$, we have,
  \begin{align}
 f^{* }_{i}(\mathbf{r}) \mathbf{dr} &= \mathbb{P}[ r \in \mathbf{dr} \vert i = i^{*}] \nonumber\\
 &=  \frac{\mathbb{P}[ \{r \in  \mathbf{dr} \}\cap \{i = i^{*}]\}}{\mathbb{P}[  i = i^{*}]} \nonumber \\
 & \stackrel{(a)}{=}  \frac{\mathbb{P}[ \{r \in  \mathbf{dr} \}\cap_{j \neq i} \{ \pi_j(\mathbf{r}_j,\lambda_j) \leq \pi_i(\mathbf{r}, \lambda_j)  ]\}}{\mathbb{P}[  i = i^{*}]} \nonumber  \\
 & \stackrel{(b)}{=} f_i(\mathbf{r})\prod_{j=1,j \neq i}^{T} F_{\pi_j}( \pi_i(\mathbf{r},\lambda_i)) \frac{1}{p_{i}} \mathbf{dr},
 \end{align}
 where $p_{i}$ is the probability that $i = i^{*}$ and $\mathbf{dr}$ is an infinitesimal element of $\mathbb{R}^{L}$. Here $(a)$ follows from the definition of $i^{*}$ in \eqref{eqn:association_rule} and $(b)$ follows from the independence of the different point process and as a consequence independence of the observation vectors $\mathbf{r}_j$.
 %Independence of $y_i$ across $i$ yields that $p_{i} = \int_{y \in \mathbf{R}} q_i(y) \prod_{j=1,j\neq i}^{T} Q_j(y) dy$. Note that $\sum_{i=1}^{T}p_i = 1$. The above calculations give us that
% \begin{align}
% w^{* }_{i}(\mathbf{r})  = w_i(\mathbf{r})\prod_{j=1,j \neq i}^{T} Q_j( f(\mathbf{r},\lambda_i)) \frac{1}{p_{i}}
% \label{eqn:defn_wistar}
% \end{align}
 \end{IEEEproof}
 \subsection{Proof of Theorem \ref{thm:cover_prob}}
 
\begin{IEEEproof} The performance of a policy $\pi_i(\cdot)$ in \eqref{eqn:perf_metric} becomes:
\begin{align}
& \mathcal{R}^{\pi}_{I}  = \mathbb{E}[p_{i^{*}}(\text{SINR}_{0}^{i^*,j_{i^*}})]   \nonumber\\
& =\sum_{i=1}^{T} \mathbb{P}[i = i^{*}] \mathbb{E} [p_i(\text{SINR}_{0}^{i,j_i})\vert  i = i^{*}] \nonumber \\
& \stackrel{}{=} \sum_{i=1}^{T} \mathbb{P}[i = i^{*}] \mathbb{E}_{\mathbf{r}_i}[\mathbb{E} [p_i(\text{SINR}_{0}^{i,j_i})\vert \mathbf{r}_i, i = i^{*}]] \nonumber\\
& \stackrel{(a)}{=} \sum_{i=1}^{T} \mathbb{P}[i = i^{*}]  \int_{\mathbf{r} \in \mathbb{R}^{L}} \mathbb{E} [p_i(\text{SINR}_{0}^{i,j_i})\vert \mathbf{r}, i = i^{*}]  f^{* }_{i}(\mathbf{r}) dr \nonumber \\
& \stackrel{(b)}{=} \sum_{i=1}^{T} \mathbb{P}[i = i^{*}]  \int_{\mathbf{r} \in \mathbb{R}^{L}} \mathbb{E} [p_i(\text{SINR}_{0}^{i,j_i})\vert \mathbf{r}]  f^{* }_{i}(\mathbf{r}) dr \nonumber \\
& \stackrel{(c)}{=} \sum_{i=1}^{T}   \int_{\mathbf{r} \in \mathbb{R}^{L}} \mathbb{E} [p_i(\text{SINR}_{0}^{i,j_i})\vert \mathbf{r}]  f_i(\mathbf{r})\prod_{j=1,j \neq i}^{T} F_{\pi_j}( \pi_i(\mathbf{r},\lambda_i)) dr. \nonumber
\end{align}
 We use the definition of $f_{i}^{*}(\mathbf{r})$ to perform the averaging over $\mathbf{r}_i$ on the event $i=i^*$ in step $(a)$. Step $(b)$ follows from the independence of $\phi_i$ across $i$ and hence we can drop the conditioning on $i = i^*$. Step $(c)$ follows from Lemma \ref{lem:defn_wistar}.
\end{IEEEproof} 
 
\subsection{Proof of Theorem \ref{thm:coverage}}
\begin{IEEEproof}
Consider first the case with information $r_1^i$:
\begin{align}
&c_{p}(1;r,\lambda,P,N_0,\beta) \nonumber \\ &{=} \mathbf{P}\left[ \frac{Ph_{1}r^{-\alpha}}{N_0 + \sum_{j\geq2} Ph_{j}r_{j}^{\alpha}} > \beta \big{\vert} r_1=r\right] \nonumber \\
& \stackrel{(a)}{=} \exp \left( -\mu \beta N_0 P^{-1}r^{\alpha}\right) \mathbb{E}\left[\exp \left(-\mu \beta r^{\alpha} \sum_{j \geq 2} r_{j}^{-\alpha}h_j \right)\right] \nonumber ,\\
\end{align}
where $(a)$ follows follow from the fact that $\{h_i\}_{\{i \geq 1\}}$ are i.i.d. exponential random variables with mean $\frac{1}{\mu}$. Simplifying $\mathbb{E}\left[\exp \left(-\mu \beta r^{\alpha} \sum_{j \geq 2} r_{j}^{-\alpha}h_j \right)\right] $, we get
\begin{align}
&\mathbb{E}\left[\exp \left(-\mu \beta r^{\alpha} \sum_{j \geq 2} r_{j}^{-\alpha}h_j \right)\right] \stackrel{(b)}{=} \\  & \exp \left( -2 \pi \lambda \int_{u \geq r} \left(1 - \mathbb{E}_h[e^{-h\mu \beta (\frac{u}{r})^{-\alpha}}] \right) u du\right) \nonumber \\
&\stackrel{(c)}{=}   \exp \left( -2 \pi \lambda \int_{u \geq r} \left(1 - \frac{\mu}{\mu \beta (\frac{u}{r})^{-\alpha} + \mu} \right) udu \right) \nonumber\\
&=   \exp \left( -2 \pi \lambda \int_{u \geq r} \left( \frac{1}{1 + \beta^{-1}(\frac{u}{r})^{\alpha}} \right) udu \right) \nonumber ,
\end{align}
where $(c)$ follow from the fact that $\{h_i\}_{\{i \geq 1\}}$ are i.i.d. exponential random variables with mean $\frac{1}{\mu}$. $(b)$ follows from the expression for the Probability Generating Functional of an independently marked PPP and the fact that conditioned on the distance of the nearest point to the origin of a PPP of intensity $\lambda$ as $r_1$  , the point process on $\mathbb{R}^2 \setminus B(0,r_1)$ is a homogeneous PPP with intensity $\lambda$.

The second case with information $[r_1^i,r_2^i]$ can be proven similarly:
\begin{align}
&c_{p}(1;[r_1, r_2],\lambda,P,N_0,\beta) \nonumber \\&  {=} \mathbf{P}\left[ \frac{Ph_{1}r_{1}^{-\alpha}}{N_0 + Ph_2 r_{2}^{-\alpha} + \sum_{j\geq3} Ph_{j}r_{j}^{\alpha}} > \beta \big{\vert} r_1, r_2\right]  \nonumber\\
& \stackrel{(a)}{=} \exp \left( -\mu \beta N_0 P^{-1}r_{1}^{\alpha}\right) \mathbb{E} \left[ \exp \left( -\mu \beta  h_2 \left( \frac{r_{1}}{r_2}\right)^{\alpha}\right) \bigg| r_1,r_2\right] \nonumber \\ &\mathbb{E}\left[\exp \left(-\mu \beta r_{1}^{\alpha} \sum_{j \geq 3} r_{j}^{-\alpha}h_j \right)\right] \nonumber.\\
\end{align}

The computation for $\mathbb{E}\left[\exp \left(-\mu \beta r_{1}^{\alpha} \sum_{j \geq 3} r_{j}^{-\alpha}h_j \right)\right]$ follows the steps similar to the above case and we skip it for brevity. We can compute $\mathbb{E} \left[ \exp \left( -\mu \beta  h_2 \left( \frac{r_{1}}{r_2}\right)^{\alpha}\right) \bigg| r_1,r_2\right]$ since $H_2$ is an independent exponential random variable and hence that expectation is equal to $ \left( \frac{1}{1 + \beta \left(\frac{r_1}{r_2}\right)^{\alpha} }\right)$. %This concludes the Lemma.
%
%\begin{align}
%&\stackrel{(b)}{=} \exp \left( -\mu \beta N_0 P^{-1} r_{1}^{\alpha}\right)    \left( \frac{1}{1 + \beta \left(\frac{r_1}{r_2}\right)^{\alpha} }\right)     \exp \left( -2 \pi \lambda \int_{u \geq r_2} \left(1 - \mathbb{E}_h[e^{-h\mu \beta (\frac{u}{r_1})^{-\alpha}}] \right) u du\right)\nonumber\\
%&\stackrel{(c)}{=} \exp \left( -\mu \beta N_0 P^{-1} r_{1}^{\alpha}\right)      \left( \frac{1}{1 + \beta \left(\frac{r_1}{r_2}\right)^{\alpha} }\right)   \exp \left( -2 \pi \lambda \int_{u \geq r_2} \left(1 - \frac{\mu}{\mu \beta (\frac{u}{r_1})^{-\alpha} + \mu} \right) udu \right)\nonumber\\
%&= \exp \left( -\mu \beta N_0  P^{-1} r_{1}^{\alpha}\right)       \left( \frac{1}{1 + \beta \left(\frac{r_1}{r_2}\right)^{\alpha} }\right)  \exp \left( -2 \pi \lambda \int_{u \geq r_2} \left( \frac{1}{1 + \beta^{-1}(\frac{u}{r_1})^{\alpha}} \right) udu \right)\nonumber,
%\end{align}
%where $(a)$ and $(c)$ follow from the fact that $\{h_i\}_{\{i \geq 1\}}$ are i.i.d. exponential random variables with mean $\frac{1}{\mu}$. $(b)$ follows from the expression for the Probability Generating Functional of an independently marked PPP and the fact that conditioned on the distance to the nearest two point to the origin of a PPP of intensity $\lambda$ as $r_1$ and $r_2$  , the point process on $\mathbb{R}^2 \setminus B(0,r_2)$ is a homogeneous PPP with intensity $\lambda$.
\end{IEEEproof}

\subsection{Proof of Lemma \ref{lem:opt_nearest}}
 \begin{IEEEproof}
\begin{align*}
F_{\pi_i}(y) &= \mathbb{P}[ c_p(1;r,\lambda_i , P_i, N_0, \beta) \leq y]\\
&= \mathbb{P}\left[ \int_{u = r}^{\infty} \frac{1}{1 + \beta^{-1}\left(\frac{u}{r} \right)^{\alpha}} u du \geq \frac{1}{2 \pi \lambda_i} \ln \left(\frac{1}{y}\right)\right] ,
\end{align*}
where the probability is with respect to the random variable $r$ which is Rayleigh distributed with parameter $\frac{1}{\sqrt{2 \pi \lambda_i}}$. In the above expression, making a change of variables of $v = \frac{u}{r}$, we have
\begin{align}
F_{\pi_i}(y) &= \mathbb{P}\left[r^2 \int_{v = 1}^{\infty} \frac{1}{1 + \beta^{-1}\left(v \right)^{\alpha}} v dv \geq \frac{1}{2 \pi \lambda_i} \ln \left(\frac{1}{y}\right)\right] \nonumber \\
& = \mathbb{P}\left[ r \geq \sqrt{{\ln\left( \frac{1}{y}\right)\frac{1}{2 \pi \lambda_i } \frac{1}{\int_{v = 1}^{\infty} \frac{1}{1 + \beta^{-1}\left(v \right)^{\alpha}} v dv }}}\right] \nonumber \\
& = e^{-  \ln\left( \frac{1}{y}\right)\frac{1}{2 } \left(\int_{v = 1}^{\infty} \frac{1}{1 + \beta^{-1}\left(v \right)^{\alpha}} v dv \right)^{-1}}.
\label{eqn:conditional_prob_1}
\end{align}
%where
%\begin{align}
%c_i(y) = \ln\left( \frac{1}{y}\right)\frac{1}{2 } \left(\int_{v = 1}^{\infty} \frac{1}{1 + \beta^{-1}\left(v \right)^{\alpha}} v dv \right)^{-1}.
%\label{eqn:c(y)}
%\end{align}
\end{IEEEproof}

\subsection{Proof of Lemma \ref{lem:distance_ratio}}

\begin{IEEEproof}
\begin{align}
\mathbb{P}\left[ \frac{r_2^i}{r_1^i} \leq x \right] &= \mathbb{E} [ \mathbb{E}  [   \mathbf{1} ( r_2^i \leq x r_1^i  )    \vert r_1^i] ] \nonumber \\
& \stackrel{(a)}{=} \mathbb{E} [ 1 - e^{-\lambda_i \pi  (x^2-1) (r^{i}_{1})^2 }] \nonumber \\
& \stackrel{(b)}{=}  1 -  \frac{1}{ x^2 }, \label{eq:chisqaure}
\end{align}

where $(a)$ follows from the Strong Markov property of a stationary PPP which states that  conditioned on $r_1$ of a PPP $\phi$, $\phi \vert_{B(0,r_1)^c}$ is a Poisson point process with the same intensity as $\phi$. The equality in $(b)$ follows from the fact that $r_{1}^2$ of a stationary PPP of intensity $\lambda$ is an exponential random variable with mean $\frac{1}{\lambda \pi}$.
\end{IEEEproof}

\subsection{Proof of Corollary \ref{cor:max_ratio_coverage}}

\begin{IEEEproof}
In employing Theorem \ref{thm:cover_prob}, we need to compute $\mathbb{E} [ p_i(\text{SINR}_{0}^{i,j_i})  \vert \mathbf{r} ]$ as follows
\begin{align}
&\mathbb{E} [ p_i(\text{SINR}_{0}^{i,j_i})  \vert \mathbf{r} ] = \mathbb{P} \left[ \frac{P_i h_{1} l(r^{i}_{1})}{  N_0 + \sum_{z \geq 2} P_i h_z l(r^{i}_{z})   } \geq \beta \bigg| \frac{r^{i}_{2}}{r^{i}_{1}}\right] \nonumber \\
&=  \int_{u=0}^{\infty} \left( c_p\left(1; \left[u ,ut \right], \lambda_i, P_i,\beta \right) g_{r^{i}_{1} \vert \frac{r^{i}_{2}}{r^{i}_{1}} } (u,t)  du \right),
\end{align}
where $c_p\left(1; \left[u ,ut \right], \lambda_i, P_i,\beta \right) $ is computed through \eqref{eqn:lemm_2nd_dist} and the conditional pdf $g_{r^{i}_{1} \vert \frac{r^{i}_{2}}{r^{i}_{1}} } (u,t) $ is the distribution of $r^i_1$ conditioned on the ratio $\frac{r^{i}_{2}}{r^{i}_{1}} = t$. %This conditional pdf can be computed as follows
\begin{align}
g_{r^{i}_{1} \vert \frac{r^{i}_{2}}{r^{i}_{1}} } (u,t) &=   \frac{g_{r^{i}_{1} , \frac{r^{i}_{2}}{r^{i}_{1}} } (u,t)}{   \int_{u=0}^{\infty} g_{r^{i}_{1} , \frac{r^{i}_{2}}{r^{i}_{1}} } (u,t) du     }
 \stackrel{(a)}{=} \frac{g_{r^{i}_{1} , \frac{r^{i}_{2}}{r^{i}_{1}} } (u,t)}{  f_{\pi_i}(t)     } \nonumber \\
& \stackrel{(b)}{=} 2(\pi \lambda_i)^2 u^3 t^4 e^{-\lambda \pi (ut)^2},
\end{align}
where $(a)$ follows from the fact that the observation  is the ratio $\frac{r^{i}_{2}}{r^{i}_{1}}$ and hence the marginal  the pdf $f_{\pi_i}(\cdot)$, which is the derivative of $F_{\pi_i}(\cdot)$ given in Lemma \ref{lem:distance_ratio}. We now show that $(b)$ holds.

Let the function $g_{r_{1}, {r_{1}}/{r_{2}}}(x,y) = (2 \pi \lambda_i)^2xy e^{-\pi \lambda_i y^2} $ denote the joint probability density function for the distance from the origin to the nearest BS  and the ratio of distances of the nearest and the second-nearest BSs distributed as a PPP of intensity $\lambda$. Transforming this pdf through $(x,y) \rightarrow (x, \frac{y}{x})$ yields the joint pdf of  $g_{r^{i}_{1}, {r^{i}_{2}}/{r^{i}_{1}}}(u,v) = (2 \pi \lambda_i)^2 u^3v e^{-\lambda \pi (uv)^2}$.
%
%
%
%$ g_{r^{i}_{1} \vert \frac{r^{i}_{1}}{r^{i}_{2}} } (u,t) = \frac{g_{r^{i}_{1} , \frac{r^{i}_{1}}{r^{i}_{2}} } (u,t)}{   \int_{u=0}^{\infty} g_{r^{i}_{1} , \frac{r^{i}_{1}}{r^{i}_{2}} } (u,t) du     }$ can be computed from Lemma \ref{lem:cond_pdf} and $c_p\left(1; \left[u ,\frac{u}{t} \right], \lambda_i, P_i,\beta \right) $ is given in Equation \eqref{eqn:lemm_2nd_dist}.
%
Plugging the law $F_{\pi_i}(\cdot)$ of Lemma \ref{lem:distance_ratio} into \eqref{eqn:coverage_prob_thm} of Theorem \ref{thm:cover_prob} finally completes the proof.
\end{IEEEproof}

\subsection{Proof of Theorem \ref{thm:max_ratio}}
\begin{IEEEproof}
We start by rearranging \eqref{eqn:coverage_prob_thm} for our special case where the observations are scalar and the performance $p_i(\cdot)$ and association $\pi_i(\cdot)$ are the same for all technologies $i$ and are independent of $i$.
\begin{align}
\mathcal{R}^{\pi}_{I} &= \sum_{i=1}^{T} \int_{t \geq 1} \mathbb{E} \left[p(\text{SINR}_{0}^{i,1}) \bigg|\frac{r_2^i}{r^i_1} = t \right] f_{\pi_i}(t) \prod_{j \neq i} F_{\pi_j}(t) dt \nonumber \\
& = \sum_{i=1}^{T} \int_{t \geq 1} \mathbb{P} \left[\text{SINR}_{0}^{i,1} \geq \beta_i \bigg|\frac{r_2^i}{r^i_1} = t \right] f_{\pi_i}(t) \prod_{j \neq i} F_{\pi_j}(t) dt \nonumber \\
& \stackrel{(a)}{=}  \sum_{i=1}^{T} \mathbb{P} \left[ \text{\text{SINR}}_{0}^{i,1} \geq \beta_i, \frac{r_2^i}{r^i_1} \geq \max_{j \neq i} \frac{r_2^j}{r^j_1} \right] \nonumber \\
&= \sum_{i=1}^{T} \int_{t \geq 1} \mathbb{E} \left[ \mathbb{P} \left[ \text{\text{SINR}}_{0}^{i,1} \geq \beta_i, \frac{r_2^i}{r^i_1} \geq t \bigg|  \max_{j \neq i} \frac{r_2^j}{r^j_1} =t \right] \right] \nonumber
\end{align}
where in step $(a)$ we used the fact that the observations from the different technologies are independent.
Now using the density function of the maximum of $T-1$ independent scalar observations each distributed according to a law as given in Lemma \ref{lem:distance_ratio}, we can simplify the above equation to obtain
\begin{multline*}
\mathcal{R}^{\pi}_{I} = \left\{ \begin{array}{ll} \sum_{i=1}^{T} \int_{x \in [0,1]} \mathbb{P} [ \text{\text{SINR}}_{0}^{i,1} \geq \beta_i, \frac{r_2^i}{r^i_1} \geq \frac{1}{x} ] \\ 2(T-1)x(1-x^2)^{T-2}dx,  \mbox{ if } T \geq 2 \\ \int_{x \in [0,1]} \mathbb{P} \left[ \text{\text{SINR}}_{0}^{i,1} \geq \beta_i, \frac{r_2^i}{r^i_1} \geq \frac{1}{x} \right] dx, & \mbox{if } T=1 \end{array} \right.
\end{multline*}
Further notice that
\begin{align}
&= \mathbb{P} \left[ \text{\text{SINR}}_{0}^{i,1} \geq \beta_i, \frac{r_2^i}{r^i_1} \geq \frac{1}{x} \right] \nonumber \\
&= \mathbb{P} \left[ \text{\text{SINR}}_{0}^{i,j_i} \geq \beta_i \bigg| \frac{r_2^i}{r^i_1} \geq \frac{1}{x} \right] \mathbb{P} \left [ \frac{r_2^i}{r^i_1} \geq \frac{1}{x}\right] \nonumber \\
&= \mathbb{P} \left[ \text{\text{SINR}}_{0}^{i,1} \geq \beta_i \bigg| \frac{r_2^i}{r^i_1} \geq \frac{1}{x} \right]  x^2, \label{eqn:ratio_corr1}
\end{align}
where \eqref{eqn:ratio_corr1} follows from Lemma \ref{lem:distance_ratio}. Now it remains to show the following lemma, which finally proves \eqref{eq:max-ratio-no-noise}.
\begin{lem}
Assume $N_0^i=0$ and $l_i(r)=r^{-\alpha}$. For any technology $i$ with intensity $\lambda_i$, $$\mathbb{P} \left[ \text{\text{SIR}}_{0}^{i,1} \geq \beta_i \bigg| \frac{r_2^i}{r^i_1} \geq \frac{1}{x} \right] =  \frac{1}{1 + \beta_i^{2/\alpha} \phi(\alpha,\beta_i,x)} . $$
\label{lem:coverage3}
\end{lem}

In the meantime, akin to the derivation from \eqref{eqn:coverage_original} to \eqref{eqn:rate_cover_from_cover_prob}, we can obtain the following average achievable rate expression:
\begin{align}
& \mathbb{E} \left[ \log_2 \left( 1+ \text{\text{SIR}}_{0}^{i,1}  \right)  \bigg| \frac{r_2^i}{r^i_1} \geq \frac{1}{x} \right] \nonumber \\
&= \int_{t \geq 0} \frac{1}{1 + (2^t-1)^{2/\alpha} \phi(\alpha,2^t-1,x)} dt. \nonumber
\end{align}
After manipulating this equation in a similar way, we can derive \eqref{eq:max-ratio-rate-no-noise}.
\end{IEEEproof}

\subsection{Proof of Lemma \ref{lem:coverage3}}
\label{appendix_lemma_ref}
\begin{IEEEproof}
We drop the subscripts and superscripts $i$ denoting technologies for brevity.
\begin{align*}
&\mathbb{P} \left[  H_1 (r_1)^{-\alpha} \geq \beta \sum_{j \geq 2} H_j (r_j)^{-\alpha} \bigg| \frac{r_2}{r_1} \geq \frac{1}{x}  \right] \nonumber \\ &= \mathbb{E} \left[ \mathbb{P} \left[  H_1 (r_1)^{-\alpha} \geq \beta \sum_{j \geq 2} H_j (r_j)^{-\alpha} \bigg| \frac{r_2}{r_1} \geq \frac{1}{x}, r_1  \right] \right] \nonumber \\
&= \int_{u=0}^{\infty} \mathbb{P} \left[  H_1 (u)^{-\alpha} \geq \beta \sum_{j \geq 2} H_j (r_j)^{-\alpha} \bigg| \frac{r_2}{r_1} \geq \frac{1}{x}, r_1 = u \right] \\ &f_{r_1 \vert \frac{r_2}{r_1} \geq \frac{1}{x}}(u) du \nonumber \\
& \stackrel{(a)}{=}   \int_{u \geq 0 }\mathbb{E} \left[  \prod_{x \in \phi_i , ||x_i|| \geq u/x}   \frac{1}{1 + \beta u^{\alpha} ||x_i||^{-\alpha}}   \right] \\& \frac{1}{\mathbb{P}\left[ \frac{r_2}{r_1} \geq \frac{1}{x}\right]}\int_{v \geq 1/x} f_{r_1 , \frac{r_2}{r_1}}(u, v )dv \nonumber \\
& \stackrel{(b)}{=}  \int_{u \geq 0 } \exp \left( -2 \pi \lambda \int_{w \geq u/x}   \left( 1 - \frac{1}{1 + \beta u^{\alpha} w^{-\alpha}} \right) w dw  \right) \\&\frac{1}{x^2}2 \lambda \pi u e^{-\lambda \pi (u/x)^2} du \nonumber \\
& \stackrel{(c)}{=} \int_{u \geq 0 }   \exp \left( - \pi \lambda u^2 \beta^{\frac{2}{\alpha}}   \int_{b \geq \frac{\beta^{\frac{-2}{\alpha}}}{x^2}} \frac{1}{1 + b^{\frac{\alpha}{2}}} db \right) \frac{1}{x^2}2 \lambda \pi u e^{-\lambda \pi (u/x)^2} du \nonumber \\
& \stackrel{(d)}{=} \int_{u \geq 0 }   \exp \left( - \pi \lambda u^2 \beta^{\frac{2}{\alpha}}  \frac{1}{x^2} \int_{c \geq \beta^{\frac{-2}{\alpha}}} \frac{1}{1 + x^{-\alpha}c^{\frac{\alpha}{2}}} dc    \right)  \\&\frac{1}{x^2}2 \lambda \pi u e^{-\lambda \pi (u/x)^2} du \nonumber \\
&= \frac{1}{1 + \beta ^{2/\alpha} \phi(\alpha, \beta, x)} \nonumber
\end{align*}
where step $(a)$ follows from the fact that $\{H_i\}$ are i.i.d. exponential random variables as in the proof of Theorem \ref{thm:coverage}. Step $(b)$ follows from Lemmas \ref{lem:distance_ratio} and the PGFL of a PPP. Step $(c)$ follows by the substitution $\beta^{\frac{-1}{\alpha}}u^2b^2 = w$. Step $(d)$ follows by the substitution $x^2 b = c$.

\end{IEEEproof}

\end{document}